\documentclass[11pt,letterpaper]{article}
\pdfoutput=1
\usepackage{jheppub}
\usepackage{color}
\usepackage{graphicx}

\usepackage{verbatim}
\usepackage{amsmath}
\usepackage{amssymb}
\usepackage{subfig}
\usepackage{url}
\usepackage{bbold}
\usepackage{slashed}

\usepackage{multirow}
\usepackage{threeparttable}
\usepackage{paralist}

\newcommand{\be}{\begin{equation}}
\newcommand{\ee}{\end{equation}}

\newcommand{\pt}{$p_T$~}

\begin{document}
\title{Thinking outside the ROCs: Designing Decorrelated Taggers (DDT) for jet substructure}

\author[a]{James Dolen,}
\author[b]{Philip Harris,}
\author[a]{Simone Marzani,}
\author[a]{Salvatore Rappoccio,}
\author[c]{and Nhan Tran}

\affiliation[a]{University at Buffalo, The State University of New York, Buffalo, NY 14260-1500, USA}
\affiliation[b]{CERN, European Organization for Nuclear Research, Geneva, Switzerland}
\affiliation[c]{Fermi National Accelerator Laboratory (FNAL), Batavia, IL 60510, USA}

\abstract{
We explore the scale-dependence and correlations of jet substructure observables to improve
upon existing techniques in the identification of highly Lorentz-boosted
objects. Modified observables are designed to remove
correlations from existing theoretically well-understood observables,
providing practical advantages for experimental measurements and
searches for new phenomena.  We study such observables in $W$ jet tagging and 
provide recommendations for observables based on considerations beyond 
signal and background efficiencies.
}

\preprint{
\begin{flushright}
FERMILAB-PUB-16-046-PPD
\end{flushright}
}
\maketitle

\section{Introduction}
\label{sec:introduction}

Techniques that aim to exploit the substructure of jets in order to identify highly
Lorentz-boosted objects~\cite{boost2010,boost2011,boost2012,Adams:2015hiv} have become an essential component of the LHC phenomenology toolkit. 
Several grooming and tagging algorithms, e.g.~\cite{bdrs,jhu,catop_cms,pruning1,pruning2,trimming,nsub,softdrop,Soper:2011cr,Soper:2012pb,Soper:2014rya}, have been developed, successfully tested, and are currently used in experimental analyses.
Considerable theoretical progress has also been made and theoretical calculations that describe the action of groomers and taggers on both background~\cite{Dasgupta:2013ihk,Dasgupta:2013via} and signal jets~\cite{Feige:2012vc,Dasgupta:2015yua} have been performed. More recently, calculations have been extended to interesting case in which a jet shape is measured in conjunction with a cut on the jet mass in ~\cite{Larkoski:2014tva,Larkoski:2014gra,Larkoski:2014zma,Larkoski:2015kga} and~\cite{Dasgupta:2015lxh}.

Despite this enormous amount of progress, experimental collaborations have yet to fully
exploit these advantages to reduce systematic uncertainties in analyses using substructure techniques. 
Much study has been focused on the relationship of numerous identification observables
in order to construct the most optimal heavy object taggers.  
Dedicated phenomenological studies~\cite{Adams:2015hiv} and detailed analysis by CMS~\cite{Khachatryan:2014vla,CMS:2014fya,CMS:2014joa,CMS-PAS-JME-15-002} and ATLAS~\cite{Aad:2013gja,Aad:2015rpa,ATL-PHYS-PUB-2015-053,ATL-PHYS-PUB-2015-033} employing multivariate techniques were performed
in order to understand how to best identify boosted $W/Z$ bosons, top quarks and Higgs bosons optimizing the statistical discrimination power of background rejection and signal efficiency.
Moreover, there has been recent interest in using computer vision techniques to combine individual calorimeter cells into non-linear optimal observables~\cite{Cogan:2014oua,Almeida:2015jua,deOliveira:2015xxd}. However, a quantitative study of the reduction of systematic uncertainties by taking advantage of theoretical improvements has not yet been performed.

In the following study, we aim to build a tagger based not only on statistical discrimination power, but also the robust behavior of the inherent QCD background.  
This tagger will be designed such that, after applying a flat cut on the tagging variable, the shape of the QCD background jet mass distribution remains stable and flat. 
We demonstrate our methodology, entitled ``designed decorrelated taggers (DDT)", by performing an example analysis in which hadronically decaying $W$ boson jets are distinguished from quark- and gluon-initiated jets. The DDT approach is applicable to the identification of any heavy boosted objects, such as Z, H, and top jets.

\subsection*{Samples}
\label{sec:samples}

The Monte Carlo samples used in this study were originally used for studies in the BOOST13 report~\cite{Adams:2015hiv}.
Samples were generated at $\sqrt{s}$ = 8~TeV for QCD dijets, and for $W^+W^-$ pairs produced in the decay of a scalar resonance. 
The QCD events were split into subsamples of $gg$ and $q\bar{q}$ events, allowing for tests of discrimination of hadronic $W$ bosons, quarks, and gluons.
QCD samples were produced at leading order (LO) using {\tt MADGRAPH5}~\cite{MG5}, while $WW$ samples were generated using the {\tt JHU GENERATOR}~\cite{Gao:2010qx}. 
The samples were then showered through {\tt PYTHIA8} (version 8.176)~\cite{Sjostrand:2007gs} using the default {\tt tune 4C}~\cite{Buckley:2011ms}.
The samples were produced in exclusive $p_T$ bins of width 100~GeV at the parton level.
The $p_T$ bins investigated in this report were 300-400~GeV, 500-600~GeV and 1.0-1.1~TeV. 

The stable particles in the generator-level events are clustered into jets with the anti-k$_T$ jet algorithm~\cite{antikt} 
with three different distance parameters, $R=0.4,0.8,1.2$, using {\tt fastjet 3.1}~\cite{fastjet1,Cacciari:2011ma}. 
No multiple parton interactions (or pileup) is
used in these samples, although previous LHC measurements~\cite{jetmass_atlas,jetmass_cms} have shown that grooming algorithms 
are more resilient to pileup effects than standard jet algorithms. Furthermore, it was shown in those measurements that the
Monte Carlo simulation can accurately reproduce the data for regions of high jet mass, whereas there are disagreements below the
Sudakov peak. The grooming algorithms, however, mitigate this disagreement very strongly as well. As such, we study jets with
a grooming algorithm applied. The algorithms we have investigated are the ``modified'' mass-drop tagger (mMDT)~\cite{bdrs,Dasgupta:2013ihk}
with $z_{cut} = 0.1$, jet trimming~\cite{trimming} with $R_{sub} = 0.3$ and $f_{cut} = 0.1$, jet pruning~\cite{pruning1,pruning2},
and soft drop~\cite{softdrop} with $z_\text{cut} = 0.1$ for both $\beta=1$ and $\beta = 2$ (note that the case of $\beta=0$ is equivalent to the mMDT). 
We have found that the conclusions are not strongly dependent on the groomer used, so have
used soft-drop with $\beta=0$ (mMDT) for most of our comparisons due to its smoother scaling behavior than other groomers~\cite{Dasgupta:2013ihk}.


\section{Current taggers}
\label{sec:current}

Current heavy object jet substructure taggers employed by CMS and ATLAS often cut on some number of observables directly or through some algorithm.
Take, for example, something similar to the CMS Run 1 $W$ tagger that uses simple cuts on the $N$-subjettiness ratio $\tau_2/\tau_1$~\cite{nsub} and the soft drop jet mass~\cite{softdrop}. 
In this study, we consider the $\tau_2/\tau_1$ variable where the subjet axes are chosen using the $k_T$ one-pass axes optimization technique.

\begin{figure}[thb]
  \centering
  \includegraphics[scale=0.38]{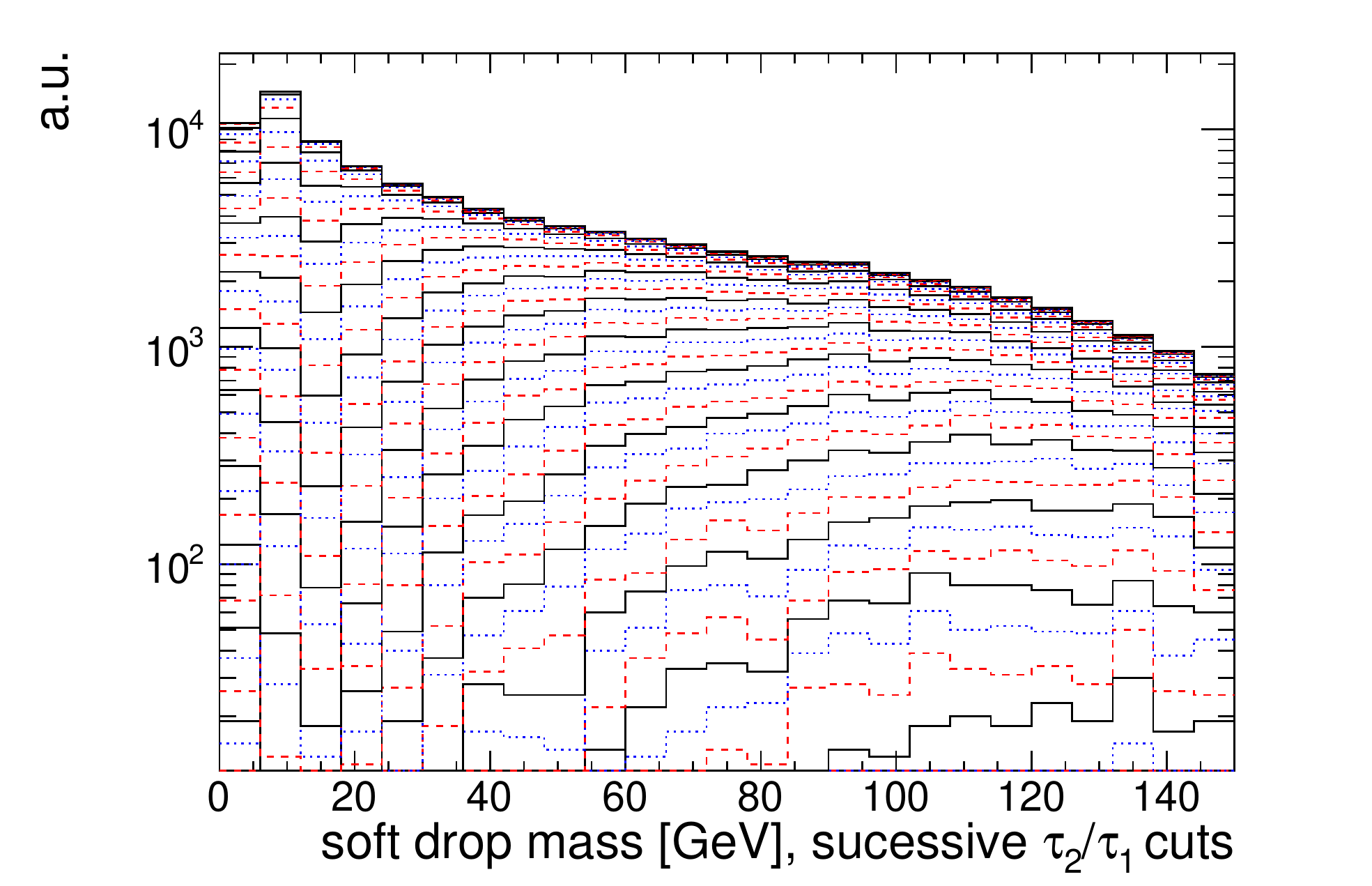}
  \includegraphics[scale=0.38]{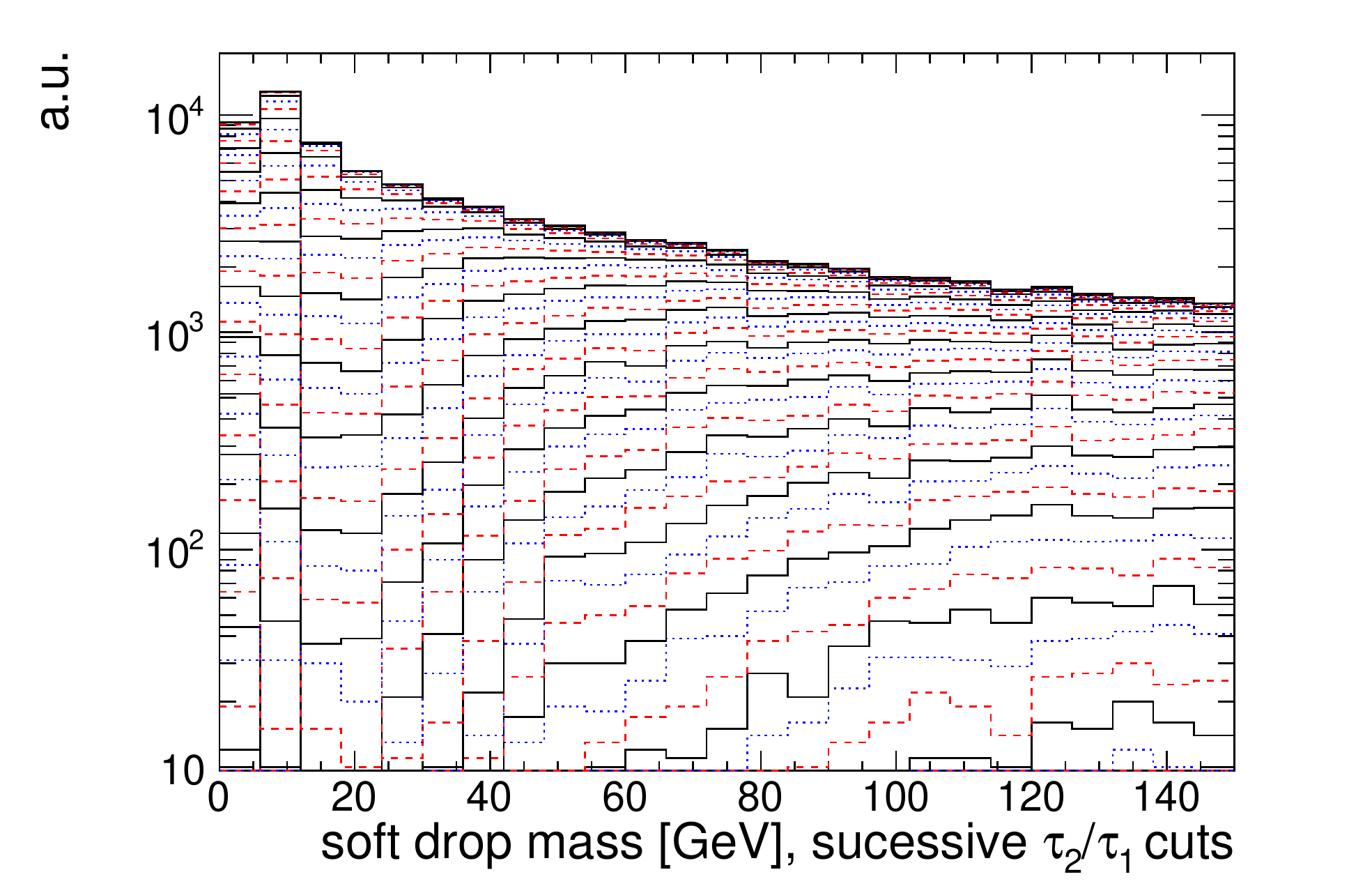}\\
\includegraphics[scale=0.38]{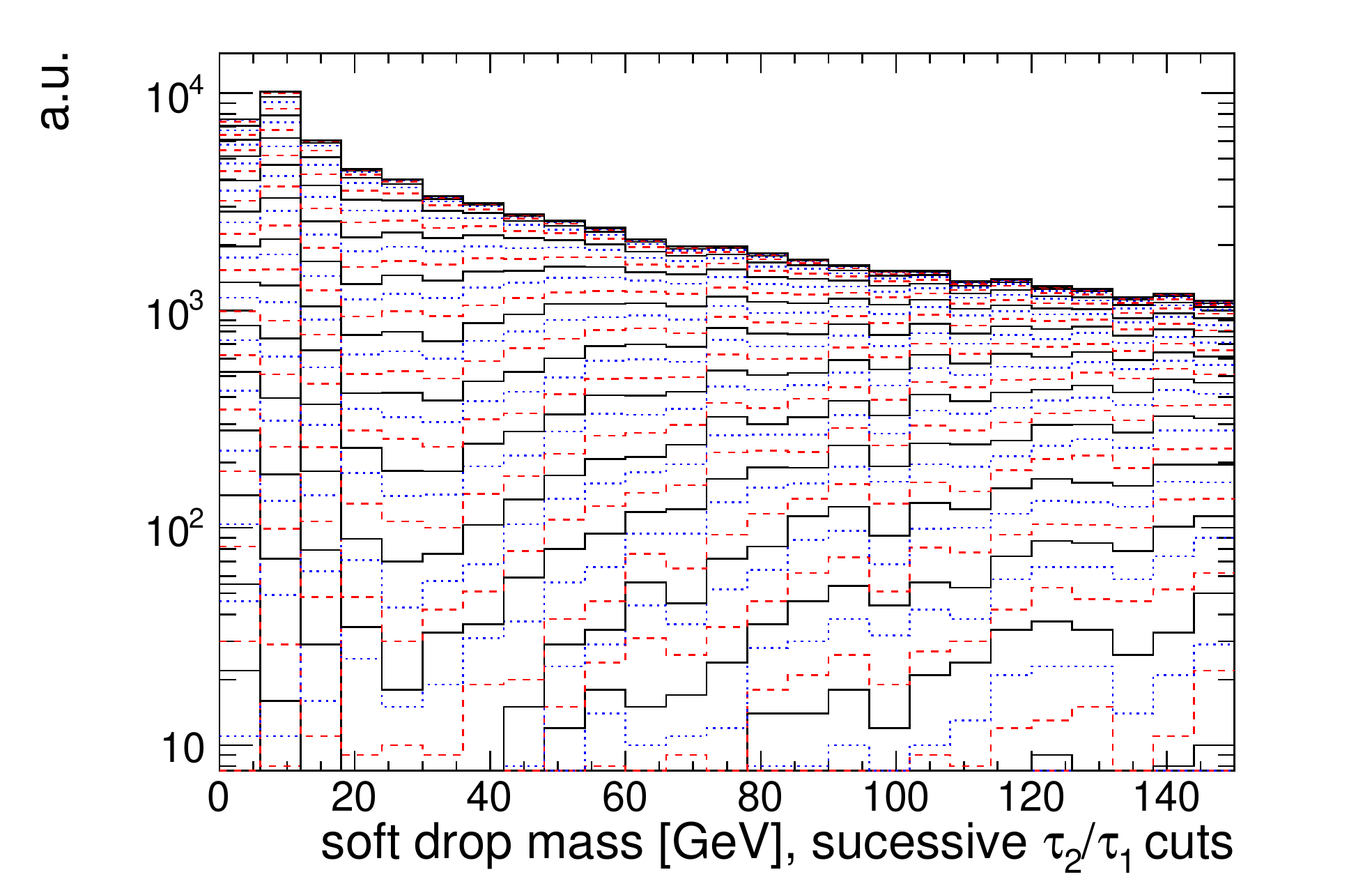}
\includegraphics[scale=0.38]{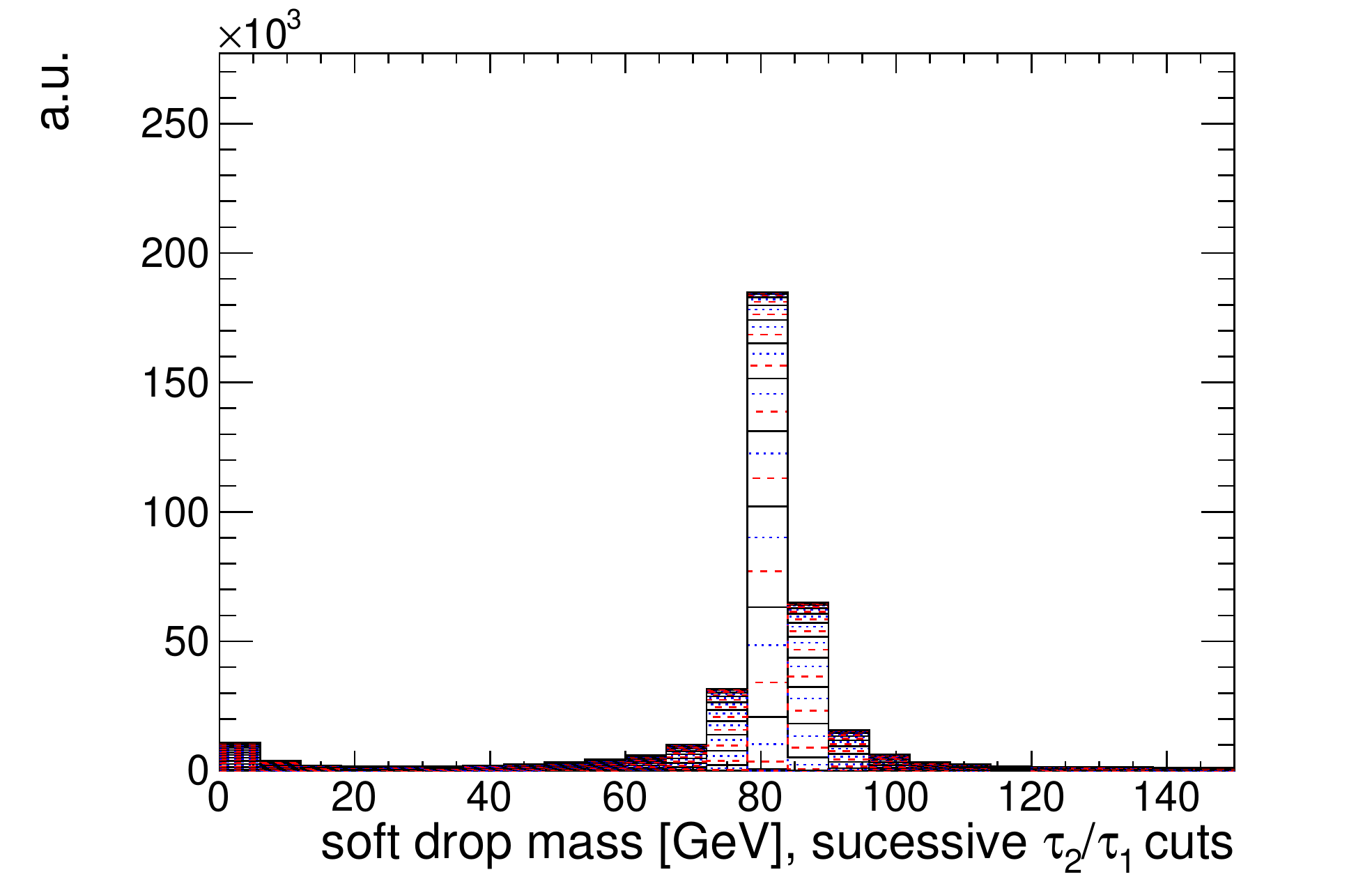}
  \caption{Soft drop mass distribution ($z_\text{cut}=0.1$ and $\beta=0$) for gluon jets after various cuts on $\tau_2/\tau_1$ ($\beta_{\tau} = 1$) for different jet $p_T$ bins: $p_T$ = 300-400~GeV (top left), 
  $p_T$ = 500-600~GeV (top right), $p_T$ = 1-1.1~TeV (bottom left) and also for the signal (bottom right), distributions for signal are stable versus $p_T$.  
  The cuts in $\tau_2/\tau_1$ vary from 1.0 to 0.0 in steps of 0.02; the changing line styles for successive cuts are meant to visually aid the reader.}
  \label{fig:msdwitht21cuts}
\end{figure}

In order to distinguish hadronically decaying $W$ bosons (which give rise to jets that are intrinsically two-pronged) from QCD background, a flat cut on
 on $\tau_2/\tau_1$ is typically performed. As expected, this procedure greatly reduces the background, but it also leads to an unwanted sculpting of the soft drop jet mass distribution (an undesirable feature also discussed in Ref.~\cite{Kasieczka:2015jma}), as shown in Fig.~\ref{fig:msdwitht21cuts}.

After cutting on $\tau_2/\tau_1$ to select jets which are two-pronged, the QCD background soft drop jet mass distribution becomes more peak-like in shape, making it harder to distinguish QCD jets from W jets which also have a peak in the jet mass distribution. 
The shape of the sculpted jet mass distribution, and the location of this artificial peak, varies for different jet $p_T$ regions.
This $p_T$ dependent sculpting of the jet mass distributions makes sideband methods of background estimation more difficult.  
In this case and in further examples, we primarily consider gluon-initiated jets though performance with quark-initiated jets is similar. 
Differences will be explored in greater detail in future studies.

In Ref.~\cite{Dasgupta:2013ihk} it was argued that flat QCD mass distributions could be obtained by tuning the value of the soft drop energy fraction threshold ($z_\text{cut}$), and optimal values for quark- and gluon-initiated jets were analytically derived. However, the presence of the $\tau_2/\tau_1$ cut makes this situation more complex and it requires reconsidering the issue.

Therefore, we propose additional criterion in determining a better tagging observable beyond pure statistical discrimination power. 
For similarly discriminant observables, we would like to find an observable which is (1) primarily uncorrelated with the groomed jet mass observable (or rather that has complementary correlations as far as discrimination is concerned)
and (2) maintains a desirable groomed mass behavior while scaling $p_T$.
Observables satisfying this criterion would, after applying a rectangular cut, still produce a flat groomed jet mass distribution. 

\section{Shape observable scaling in QCD}

We start our study of the correlations of substructure variables with the jet mass and $p_T$ by introducing the appropriate scaling variable for QCD jets:
\begin{equation} \label{eq:rho}
\rho = \log(m^2/p_T^2).
\end{equation}
Here we have differed from the typical definition of jet $\rho$ by removing the jet distance parameter $R^2$ from the denominator of the definition.
For now we keep $R=0.8$ fixed and leave this for future study.  Note that when we apply soft-drop, we take the mass in Eq.~(\ref{eq:rho}) to be computed on the constituents of the soft-drop jet, while the transverse momentum is the one of the original (ungroomed) jet.

\begin{figure}[thb]
  \centering
  \includegraphics[scale=0.38]{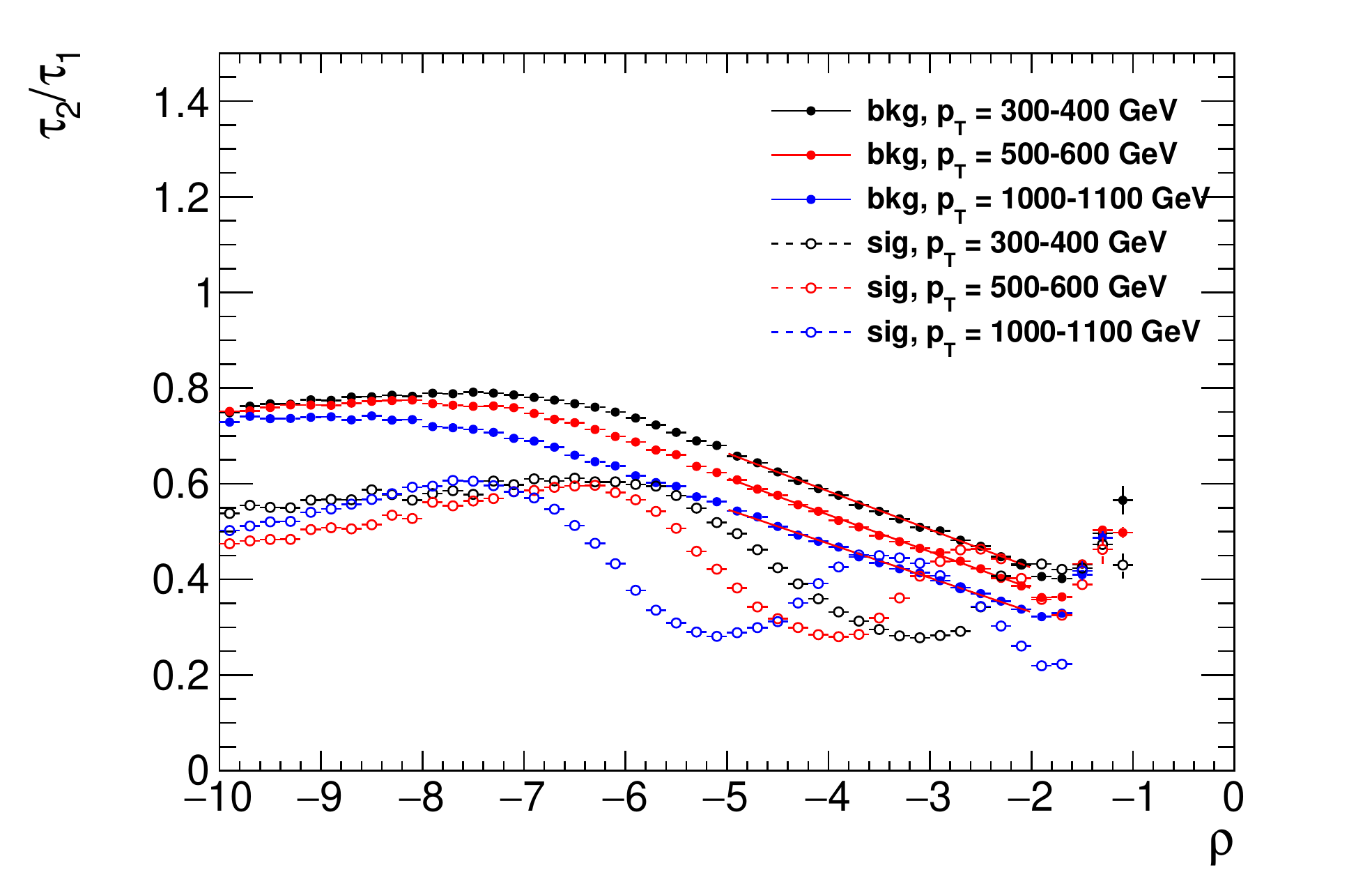}
  \includegraphics[scale=0.38]{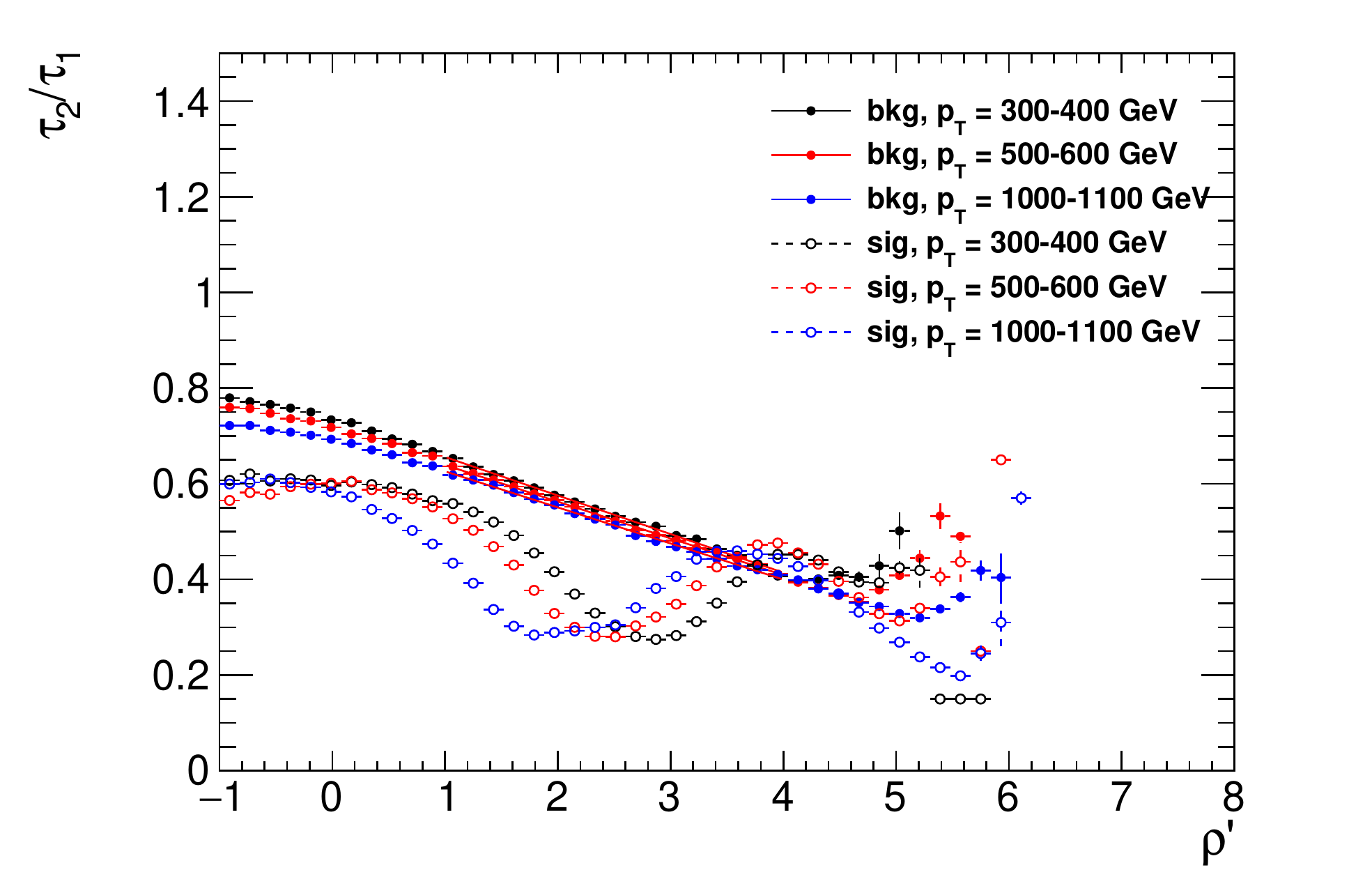}
  \caption{Profile distributions, $\langle \tau_2/\tau_1 \rangle$, as a function of $\rho = \log(m^2/p_T^2)$ (left) and as a function of $\rho^\prime = \log(m^2/p_T/\mu)$ (on the right). Solid dots correspond to background, while hollow ones to signal. The different colors correspond to different $p_T$ bins.}
  \label{fig:vsrho}
\end{figure}

We now compute, on both our background and signal samples, the average value of the $N$-subjettiness ratio $\tau_2/\tau_1$ (computed on the full jet) as a function of the soft-drop $\rho$.
This is shown in Fig.~\ref{fig:vsrho}, on the left.  
The signal W jets are shown in open circles while the background, here gluon jets, are shown in closed circles.  The various colors are different bins in jet $p_T$.
We note the typical behavior showing $\tau_2/\tau_1$ for the signal tending to lower values than the background and at a given value in $\rho$ due to the mass scale of the signal jet in a given $p_T$ bin.
The signal tends to be fixed around the W mass and thus shifts for different values of $p_T$ and is otherwise most concentrated in the dip region.
Now, let us focus on the background curves (solid points).
We notice a strong dependence on $\tau_2/\tau_1$ which is what causes the sculpting of the mass distributions shown in the previous section.  
However, we note that there exist a region in $\rho$ for which this relationship is conspicuously linear. This is an interesting behavior, which we will exploit shortly in Sec.~\ref{sec:sit}.
We also observe that, even in this linear region, there is still a residual $p_T$ dependence, which looks like, to a very good approximation, a constant shift. 
The behavior observed in Fig.~\ref{fig:vsrho} for soft drop $\rho$ is also observed for other groomers, such as trimming and pruning, within the $p_T$ ranges consdidered. At lower values of $\rho$ differences in the groomers become more apparent, most likely because in that region trimming and pruning acquire further sensitivity to soft physics~\cite{Dasgupta:2013ihk}. Thus, in the current study, we concentrate on the soft-drop mass due to its stable behavior. 

This approximate linear relation between $\tau_2/\tau_1$ can be (qualitatively) understood by noting that, in the case $\beta_\tau=2$, $\tau_2$ essentially measures the subjet mass, while $\tau_1$ corresponds to the jet mass itself. This leads to an approximately linear relation between $\tau_2 / \tau_1$ and $\rho$ in the region of the (soft-collinear) phase-space where all-order effects can be neglected.\footnote{We thank Andrew Larkoski for raising this point.}
Furthermore,  Ref.~\cite{Dasgupta:2015lxh} performed calculations for jet mass distributions in the presence of a $\tau_2/\tau_1$ cut to an accuracy which is close to next-to-leading logarithmic (NLL) accuracy. Despite the fact that the calculation corresponding to the profile plot in Fig.~\ref{fig:vsrho} were not performed, it could in principle be derived because the authors do provide the double differential distribution in $\tau_2/\tau_1$ and $\rho$.
However, some important differences between our current set-up and the one of Ref.~\cite{Dasgupta:2015lxh} prevent us from using their results to get more quantitative insight in the behaviors we observe beyond the existence of a region with linear correlation.
First Ref.~\cite{Dasgupta:2015lxh} did not consider the soft drop $\rho$ and, second, the definition of $N$-subjettiness differs in the two studies both in regards of the angular exponent ($\beta_\tau=1$ versus $\beta_\tau=2$) and of the choice of axes. 
We note that, at fixed-coupling, all the transverse momentum dependence is accounted for in the definition of the shape and $\rho$.  We have checked whether the origin of the $p_T$ dependence that we see in Fig.~\ref{fig:vsrho} (on the left) could be traced back to the transverse momentum used in the definition of the  $\rho$ (ungroomed vs groomed) but this was found not to be the case. 
Running coupling contributions, as well as other subleading corrections, do introduce a $p_T$ dependence and they are likely to responsible for the observed $p_T$ dependence. However, a quantitative understanding of these effects would require a calculation using the techniques of Ref.~\cite{Dasgupta:2015lxh}. 
This goes beyond the scope of this work and for this study we limit ourselves to a phenomenological solution, while leaving a first-principle analysis for future work. Thus, in order to remove the constant  $p_T$ dependence in the $\tau_2/\tau_1$ profile, we introduce a modified version of $\rho$:
\begin{equation} \label{eq:rhop_def}
\rho' =\rho+\log \frac{p_T}{\mu}= \log \left( \frac{m^2}{p_T \mu}\right).
\end{equation}
This change of variable, together with the choice $\mu\sim 1$~GeV, appears to perform an excellent job in getting rid of the $p_T$ dependence, as shown in Fig.~\ref{fig:vsrho}, on the right, though of course we note this is purely an empirical observation. 

\begin{figure}[thb]
  \centering
    \includegraphics[scale=0.38]{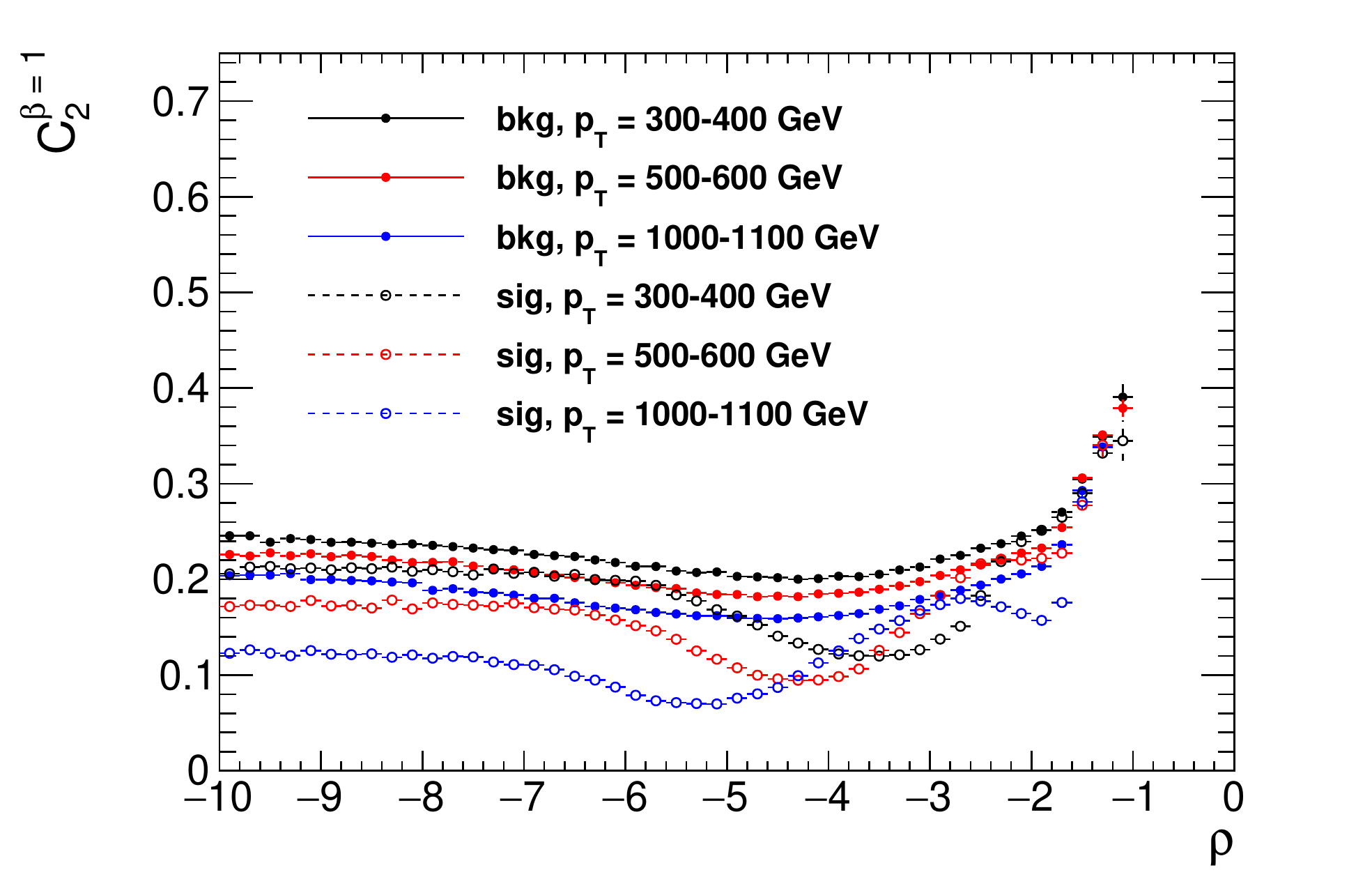}
    \includegraphics[scale=0.38]{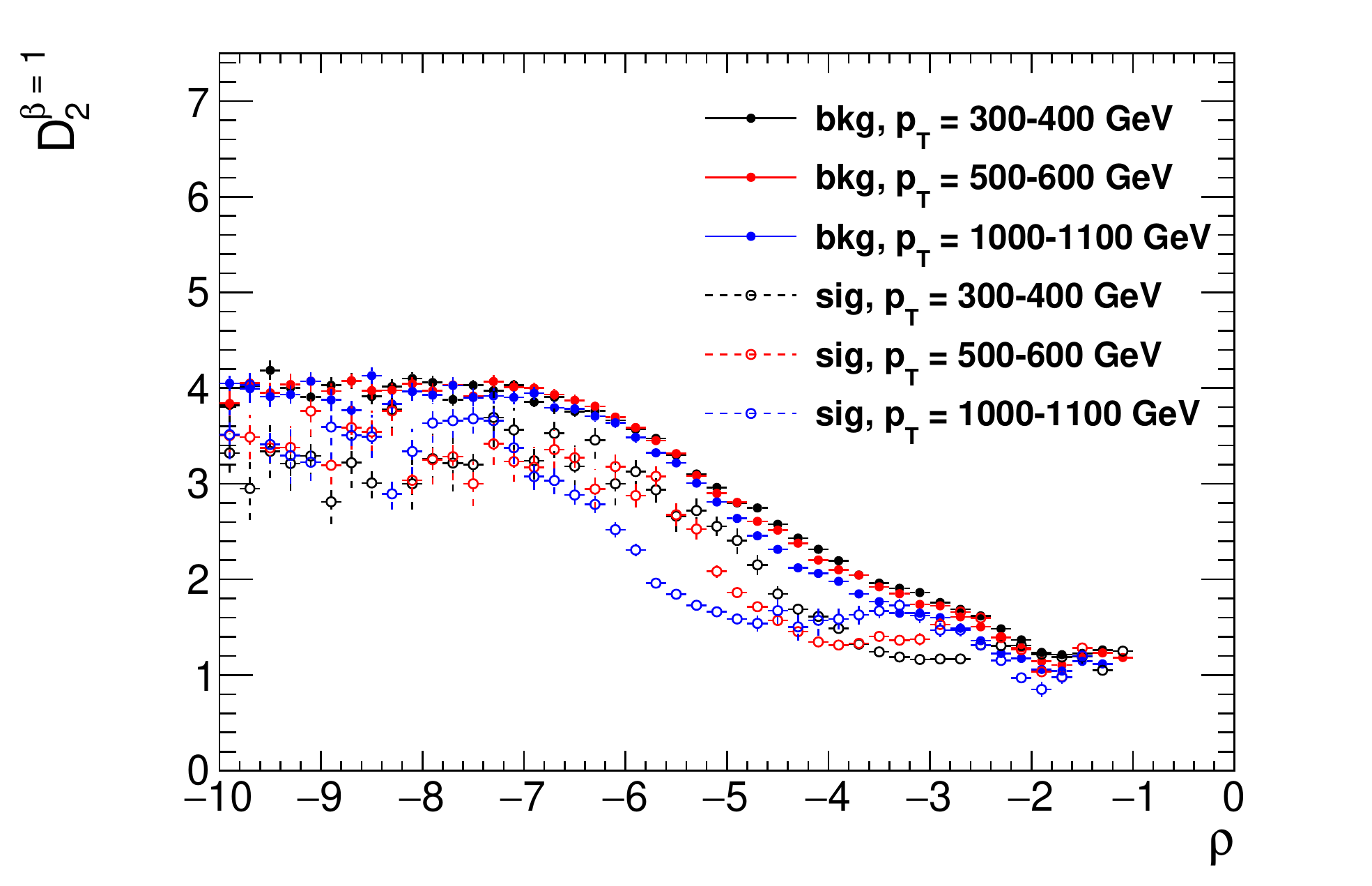}    
  \caption{Profile distributions, $\langle C_2^{\beta=1} \rangle$ (left) and $\langle D_2^{\beta=1} \rangle$ (right), as a function of $\rho = \log(m^2/p_T^2)$. Solid dots correspond to background, while hollow ones to signal. The different colors correspond to different $p_T$ bins}
  \label{fig:c2}
\end{figure}

So far, we have only considered $\tau_2/\tau_1$ versus soft drop mass.
We also noted that a similar linear correlation exists between $\tau_2/\tau_1$ and other groomed masses, though not shown explicitly.  
We can also consider other shape variables, though we leave an exhaustive exploration of all shape variables to a later study.
As an example, we show also energy correlation functions $C_2^{\beta=1}$ and $D_2^{\beta=1}$ as a function of $\rho$ in Fig.~\ref{fig:c2}.
On the left, $C_2^{\beta=1}$ shows a relatively flat distribution versus $\rho$ which is desirable although the behavior is not quite linear.
On the right, $D_2^{\beta=1}$ is highly correlated with $\rho$. 
In both cases, the correlations have some $p_T$-dependence that is not trivially empirically determined.

\section{Designing decorrelated taggers (DDT)}
\label{sec:sit}

\subsection{Transforming $\tau_2/\tau_1$ }
By performing the transformation $\rho \to \rho^\prime$, we have successfully accounted for most of the $p_T$ dependence of the profile distribution. Next we would like to perform a further transformation with the aim of flattening the profile dependence on $\rho^\prime$, with the idea that this will in turn reduce the mass-sculpting discussed earlier. 

\begin{figure}[thb]
  \centering
  \includegraphics[scale=0.38]{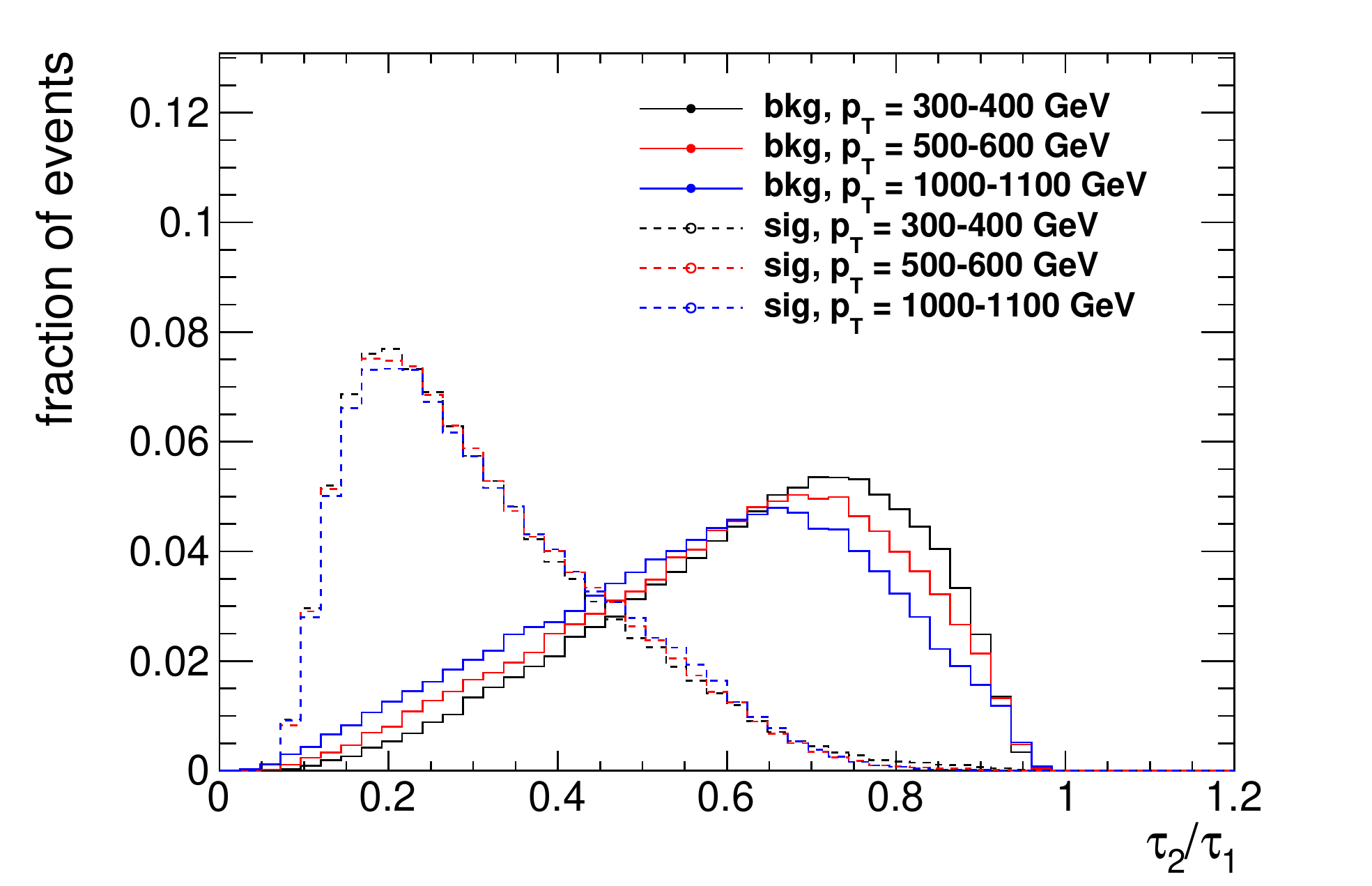}
  \includegraphics[scale=0.38]{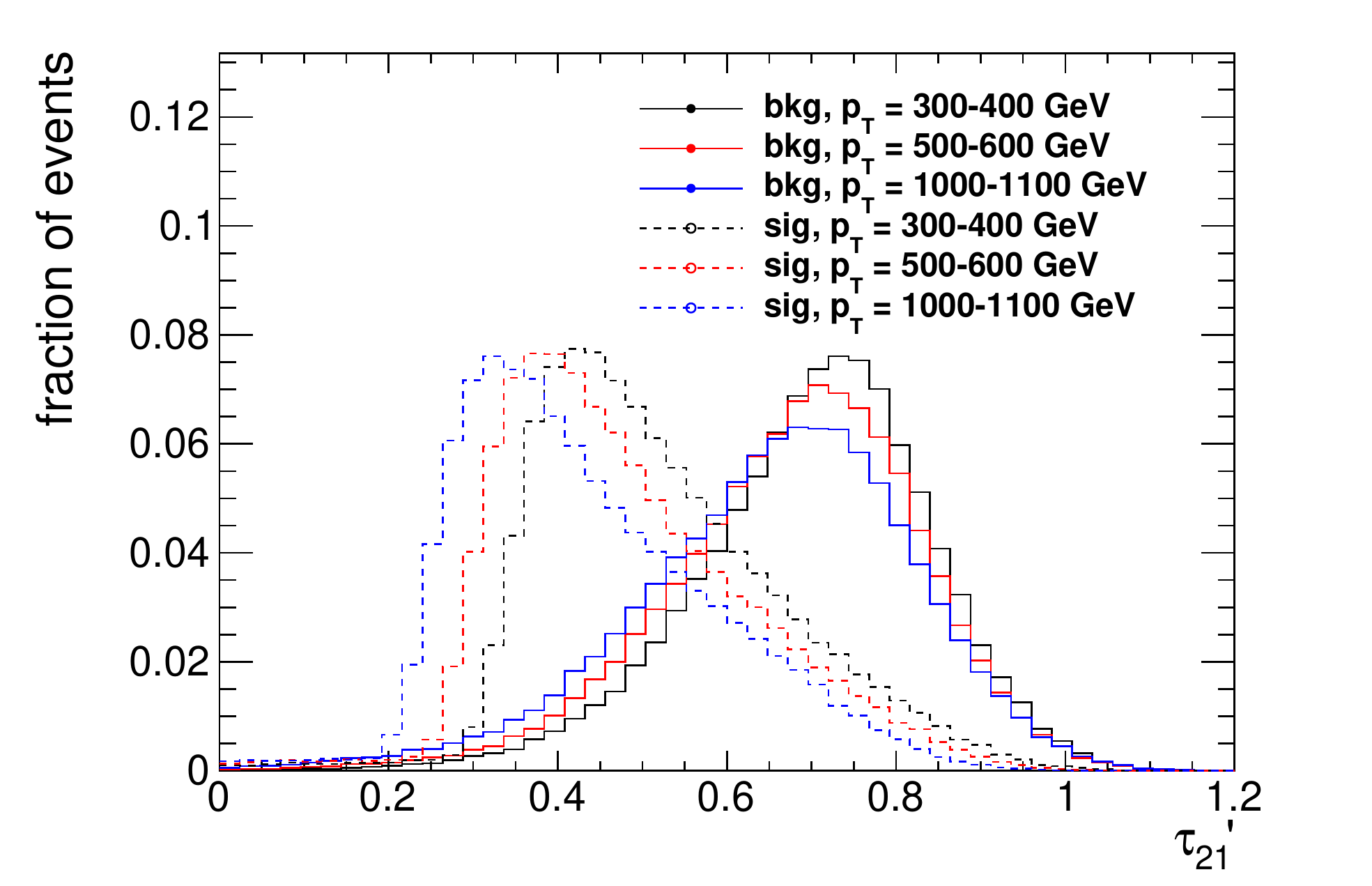}\\
  \caption{Raw $\tau_2/\tau_1$ distributions on the left and transformed distribution, $\tau_{21}^\prime$, on the right.}
  \label{fig:tau}
\end{figure}

\begin{figure}[thb]
  \centering
  \includegraphics[scale=0.50]{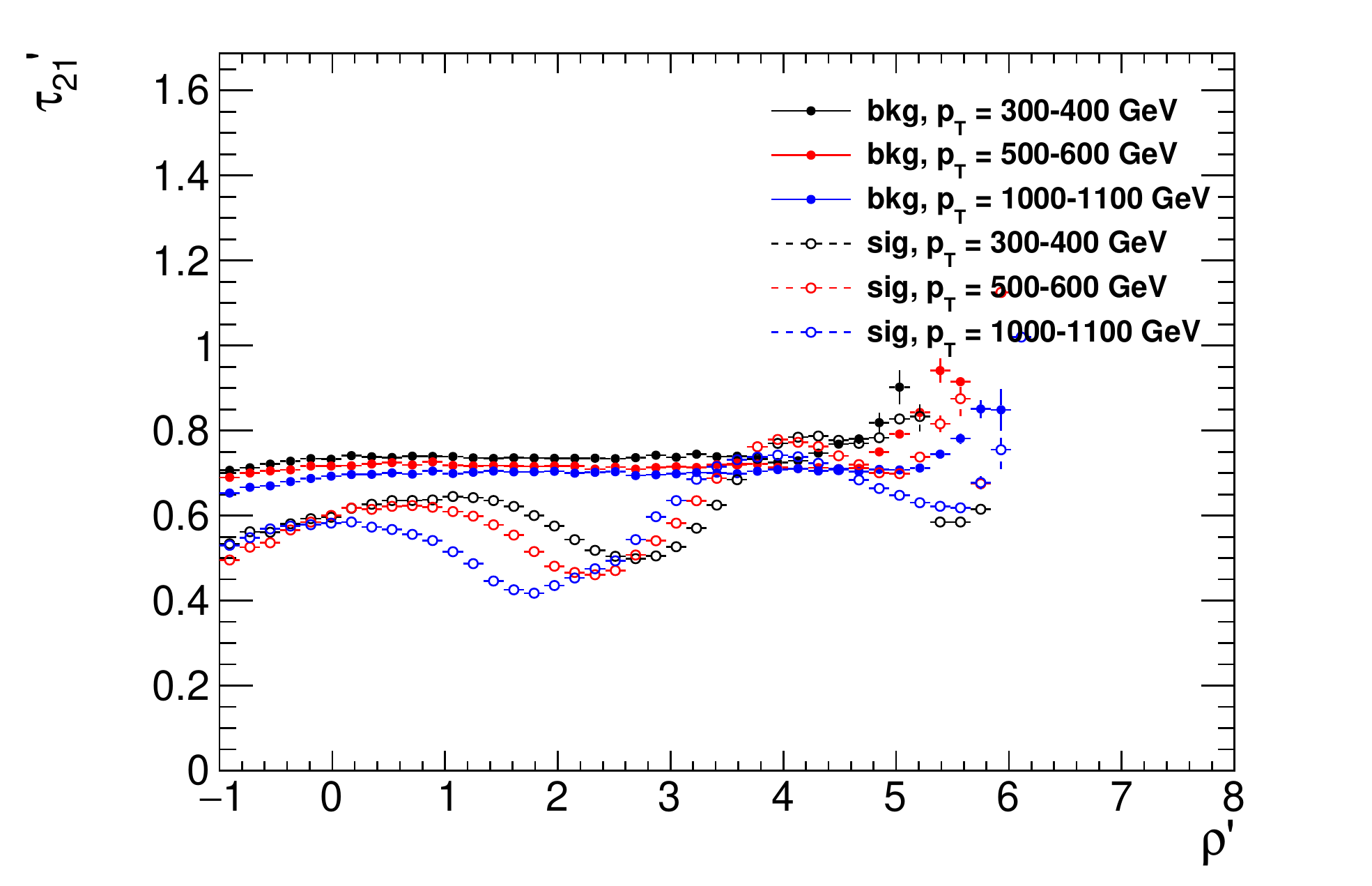}
  \caption{Profile distributions, $\langle \tau_{21}^\prime \rangle$, as a function of $\rho^\prime = \log(m^2/p_T/\mu)$. Solid dots correspond to background, while hollow ones to signal. The different colors correspond to different $p_T$ bins}
  \label{fig:vsrhoprimetrans}
\end{figure}

In order to determine the transformation we are after, we concentrate on the region in which the relationship between $\tau_2/\tau_1$ and $\rho'$ is essentially linear. Thus, we introduce 
\begin{equation}\label{eq:taup}
\tau_{21}^\prime=  \tau_2/\tau_1 - M \times \, \rho^\prime,
\end{equation}
where the slope $M$ is numerically fitted from Fig.~\ref{fig:vsrho} (red fit lines).
The comparison between the $\tau_2/\tau_1$ and $\tau_{21}^\prime$ distributions is shown in Fig.~\ref{fig:tau}, for different jet $p_T$ bins. 
The transformed variable, $\tau_{21}^\prime$, looks similar to the original variable $\tau_2/\tau_1$ although the behavior of the correlation with the groomed mass is now practically removed.  
We note that a $p_T$-dependence on the signal shape is introduced which is, in hindsight, expected given the transformation takes advantage of scaling properties of the background.  
This can cause a $p_T$-dependence in the signal efficiency with a cut on $\tau_{21}^\prime$ not present in the original $\tau_2/\tau_1$; however, we note this is not necessarily an undesirable feature.  
For example, as backgrounds decrease at higher $p_T$ it may be desirable to allow a larger signal efficiency and this should be studied in more detail in the experiments within the context of particular analyses.  
This can be seen in Fig.~\ref{fig:vsrhoprimetrans} which shows the profile of $\tau_{21}^\prime$ as a function of $\rho^\prime$ with the intended decorrelated behavior.  

\begin{figure}[thb]
  \centering
  \includegraphics[scale=0.38]{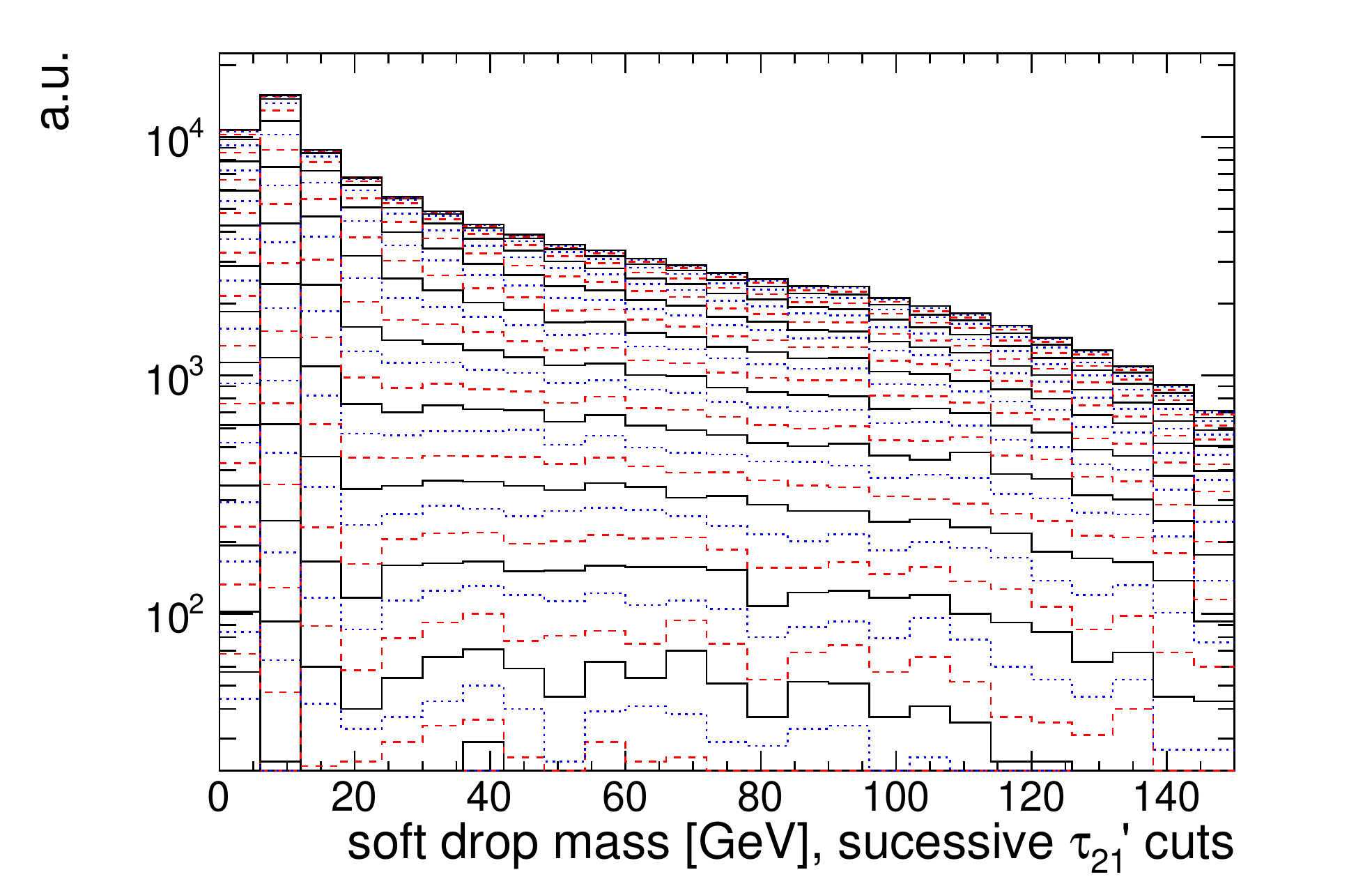}
  \includegraphics[scale=0.38]{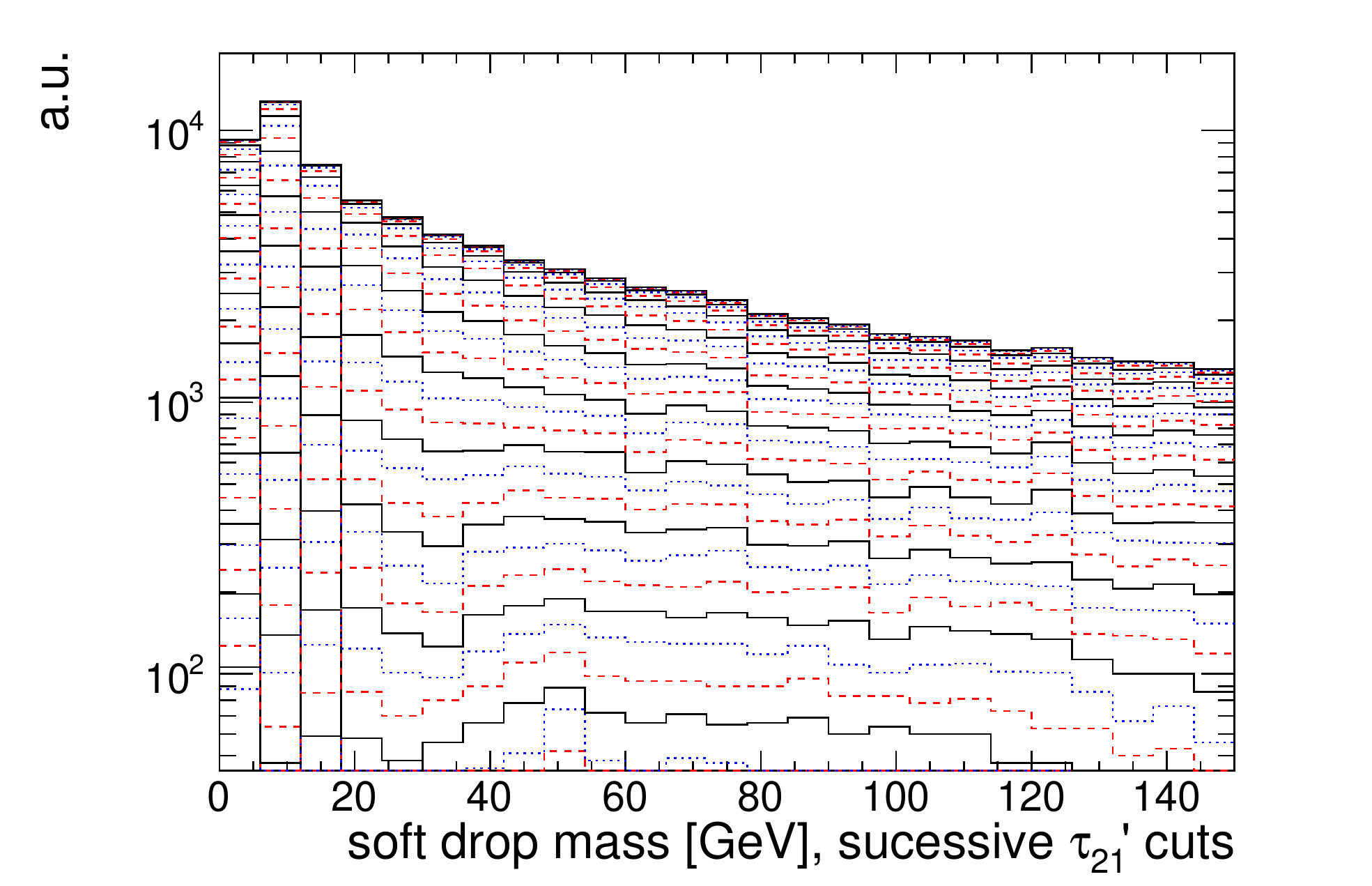}\\
  \includegraphics[scale=0.38]{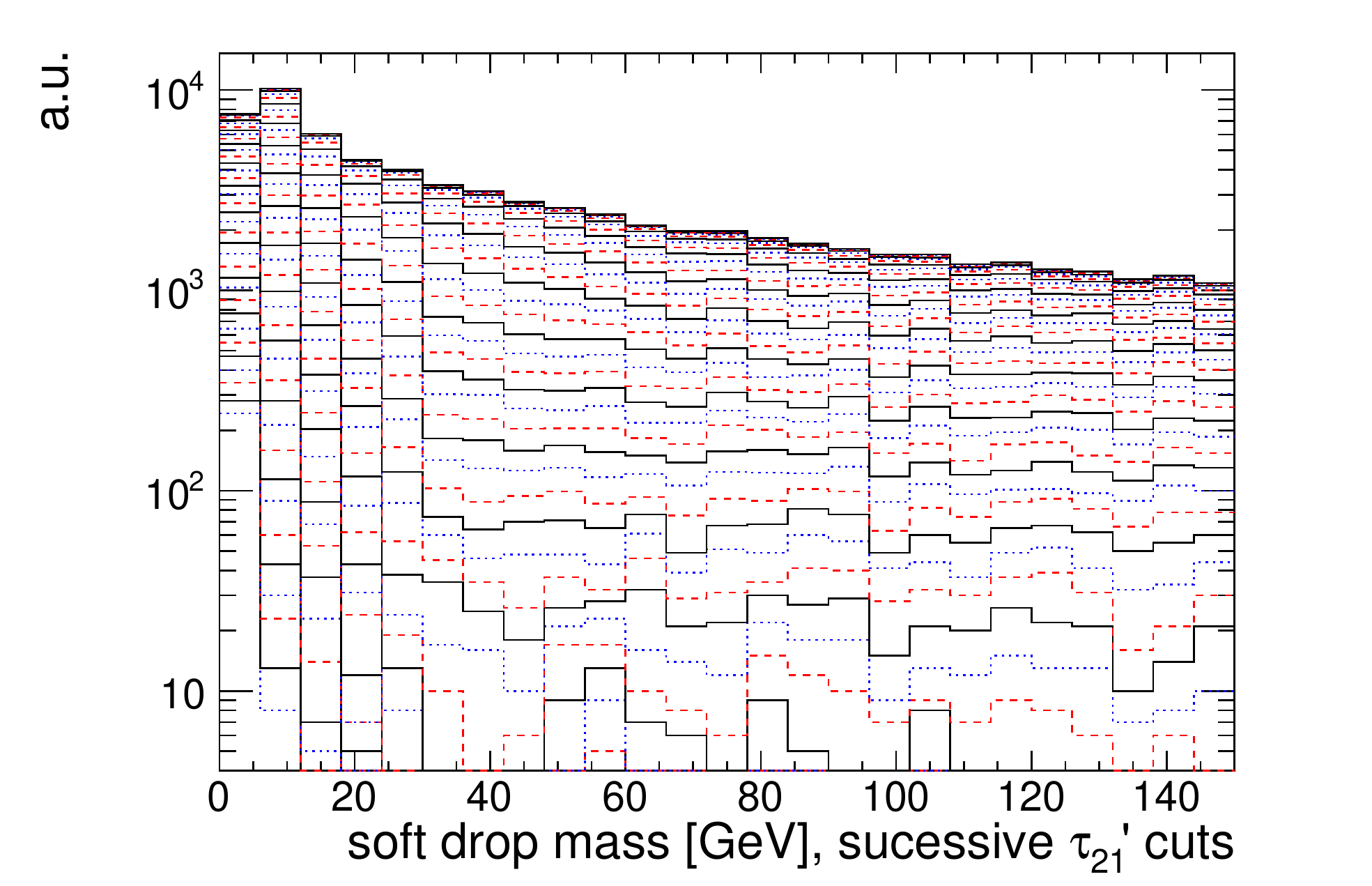}
  \includegraphics[scale=0.38]{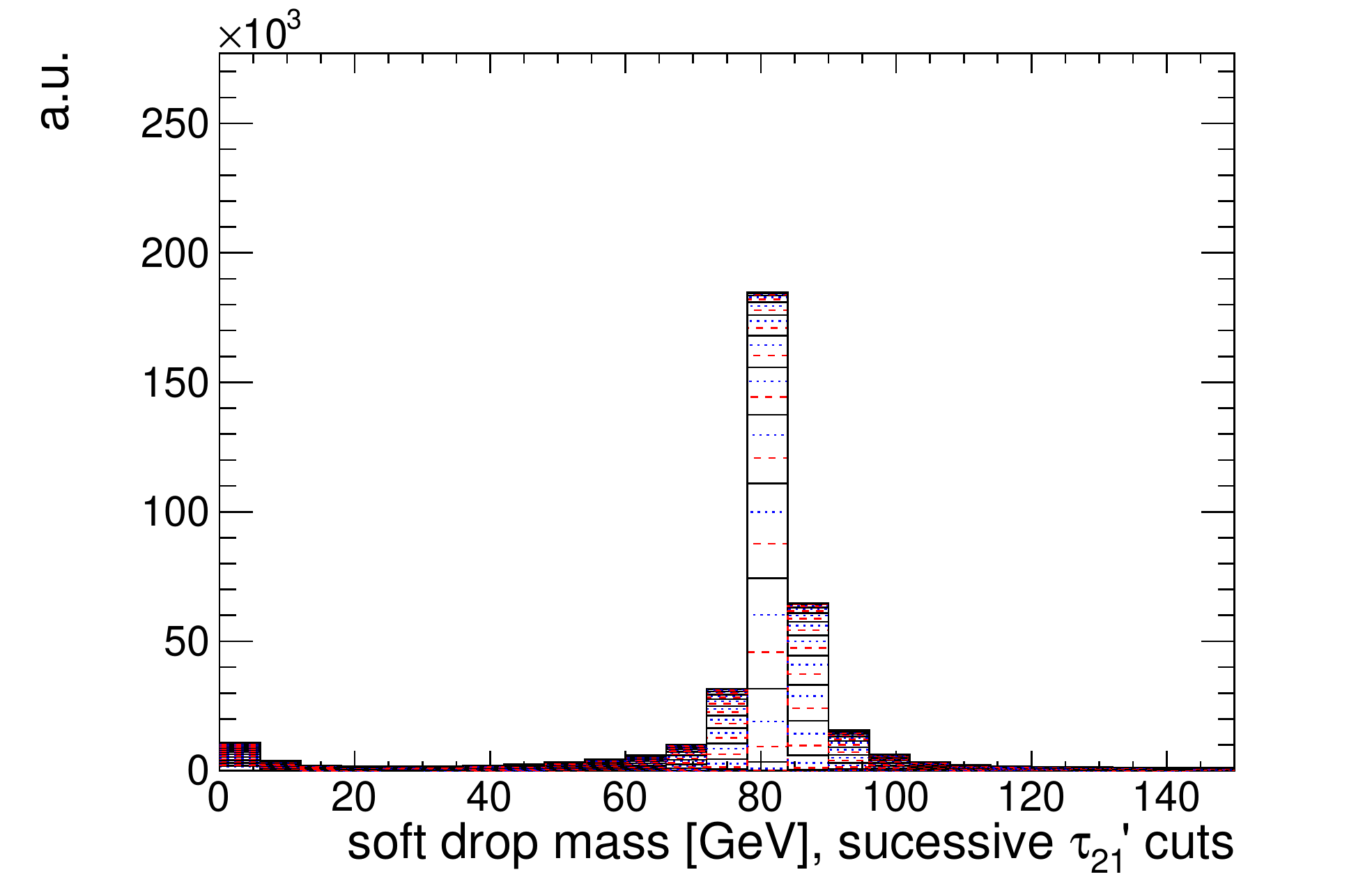}
  \caption{Soft drop mass distribution for gluon jets after various cuts on $\tau_{21}'$ for different jet $p_T$ bins: $p_T$ = 300-400~GeV (top left), 
  $p_T$ = 500-600~GeV (top right), $p_T$ = 1-1.1~TeV (bottom left) and also for the signal (bottom right), distributions for signal are stable versus $p_T$.  
  The cuts in $\tau_{21}'$ vary from 1.0 to 0.0 in steps of 0.02; the changing line styles for successive cuts are meant to visually aid the reader.}
  \label{fig:msdwitht21cutsscot}
\end{figure}

Now, we can explore the sculpting of the mass distributions making a flat cut in $\tau_{21}^\prime$. 
This is shown in Fig.~\ref{fig:msdwitht21cutsscot} which should be contrasted with Fig.~\ref{fig:msdwitht21cuts} which was obtained with a flat cut in $\tau_2/\tau_1$.  
Notice that now the sculpting of the mass distribution is considerably reduced, particularly in the region of interest where the $W$ boson peak is.
With a simple transformation, we can now preserve mass sidebands for background estimations and make robust predictions of the $p_T$ dependence of the backgrounds.  
This practical consequences of a well-behaved background shape will be explored in Section~\ref{sec:casestudies}. 
Generally speaking, a non-linear dependence is not a technical obstacle to performing an observable transformation and we discuss this in Section~\ref{ref:sec:generalized}; 
however, studying the behavior in a simple analytic regime allows us to better understand the underlying physical behavior.
The final component to evaluating the success of the observable transformation is to understand the performance of the new observable in terms of rejecting backgrounds. 

\subsection{Performance of DDT}
\label{sec:perf}

\begin{figure}[thb]
  \centering
  \includegraphics[scale=0.38]{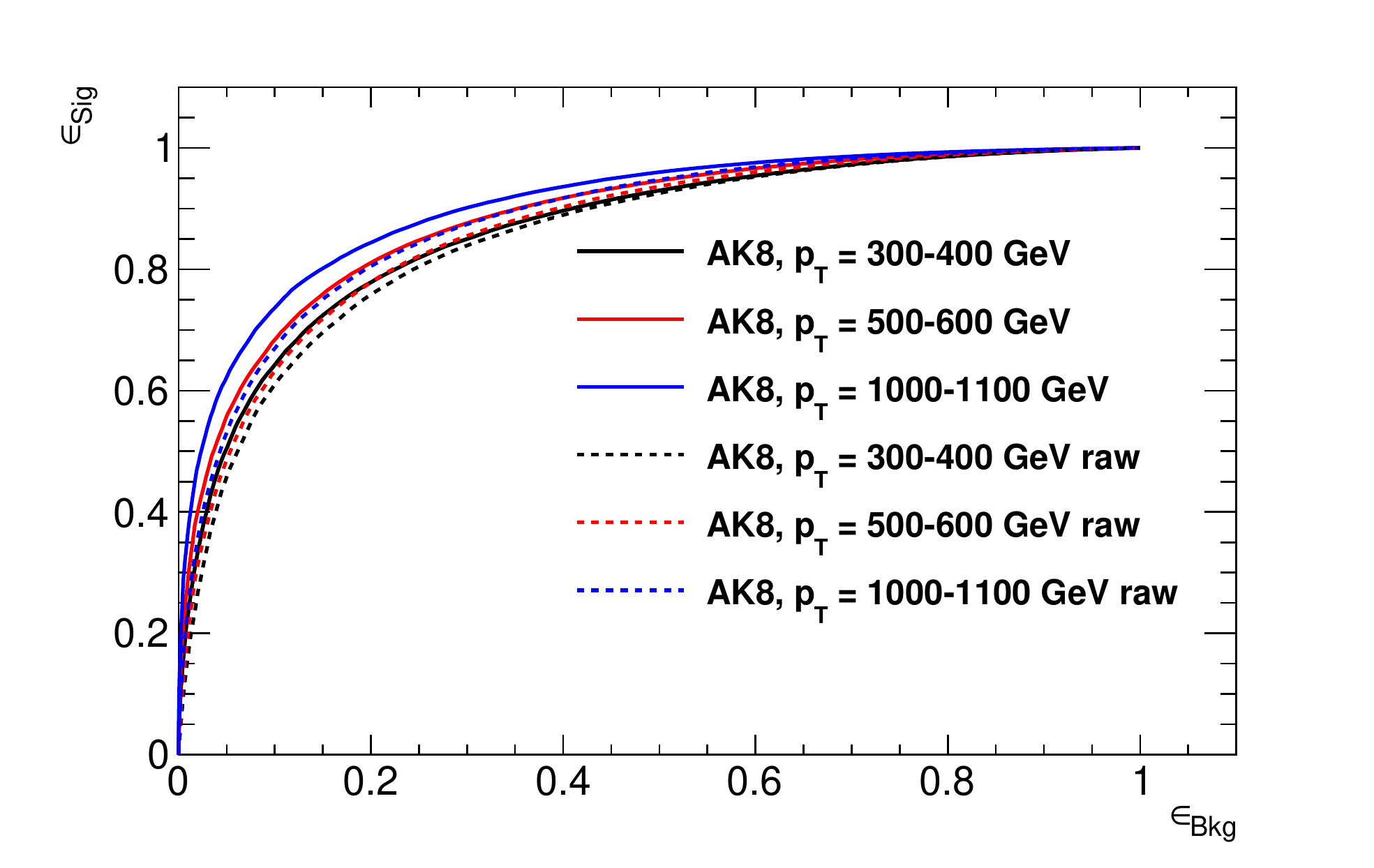}
  \includegraphics[scale=0.38]{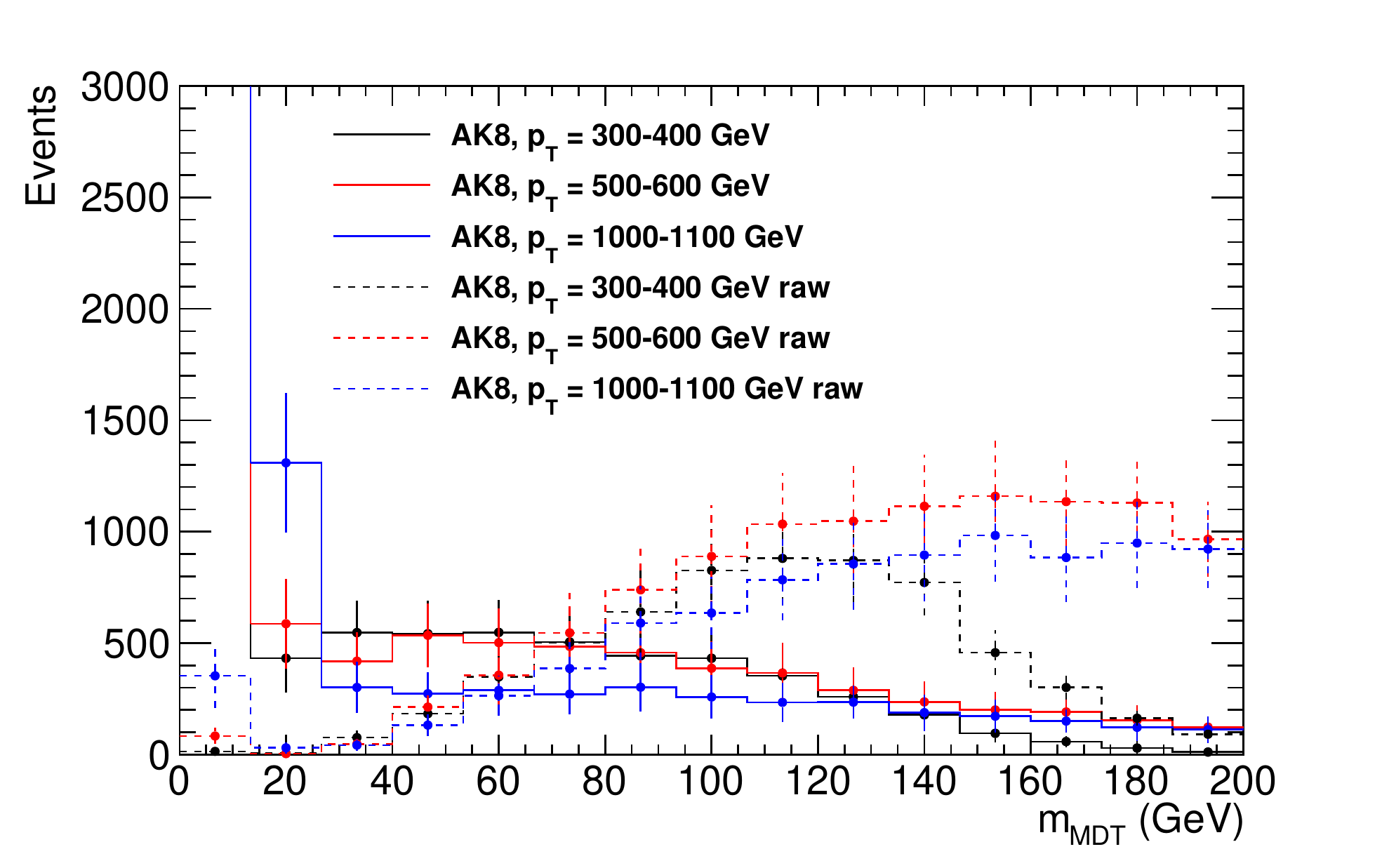}\\
  \caption{(left) ROC discrimination curve: W-tagging efficiency versus QCD jet tagging efficiency for three \pt regions
  for the transformed $\tau_{21}'$ variable (solid) and the raw $\tau_{2}/\tau_{1}$ variable (dashed). Here efficiency is defined as the number of jets with mass satisfying $60 < $mMDT$ < 120$ GeV which are tagged. (right) Soft drop mass distributions after a cut on the transformed $\tau_{21}'$ variable (solid) and the raw $\tau_{2}/\tau_{1}$ variable (dashed), where the cut corresponds to 50\% signal efficiency. Here the uncertainties on each bin signify the expected variation for a 10\% uncertainty on the W boson tag efficiency.}
  \label{fig:roc}
\end{figure}

To evaluate the performance of the transformed variable we use the traditional receiver operating characteristic (ROC) curve, defined as the signal efficiency as a function of the background efficiency.
A better discriminating tagger is characterized by higher signal efficiency and lower background efficiency.
The discriminating performance of $\tau_2/\tau_1$ and the transformed $\tau_{21}'$ are shown in the left of Fig.~\ref{fig:roc} for jets within a soft drop mass window of [60-120]~GeV (corresponding to the $W$ signal mass region).  
From the ROC curve, we note that after transforming the variable the discriminating power does not degrade and even shows modest improvement in this kinematic regime.  
We can see where this comes from in the right panel of Fig.~\ref{fig:roc}.
After cutting on raw $\tau_2/\tau_1$ the QCD soft drop jet mass distribution is sculpted such that many of the jets surviving the cut fall into the W mass region. In contrast, cutting on $\tau_{21}'$ leaves a more linearly falling distribution which preserves the low sideband.
The mass distributions on the right side of Fig.~\ref{fig:roc} are after making a cut on the shape observable to maintain a signal efficiency of 50\%.

\section{Case studies}
\label{sec:casestudies}

Currently, the systematic uncertainties in extracting the efficiency are 
large (and usually dominant) sources of uncertainty in SM and BSM analyses
at the LHC~\cite{Khachatryan:2015bma,Khachatryan:2015ywa,Khachatryan:2014gha,Khachatryan:2014hpa,Chatrchyan:2012ypy,Aad:2015owa,Aad:2015uka}. 
There are several places where the improved scaling
behavior can reduce these systematics, in addition to the performance improvements in
the ROC curves shown in Fig.~\ref{fig:roc}. We will present two
improvements, the preservation
of mass sidebands in the kinematic fit to extract the $W$ tagging
efficiency from semileptonic $t\bar{t}$ events, 
and the overall background estimate in diboson analyses. 
Both cases take advantage of the flatter background distributions to
improve the uncertainties in shape-based fits. 

\subsection{Preservation of mass sidebands}

The shape of the jet mass spectrum is used in the LHC
experiments to determine the $W$ tagging efficiency; for instance, 
CMS relies on a simultaneous fit to the jet mass
in events that pass and fail the $\tau_{21}$ selection. However, as
shown in Fig.~\ref{fig:msdwitht21cuts}, the $\tau_{21}$ selection
significantly kinematically sculpts the background distribution in
this variable. This can lead to significant fitted uncertainties when
extracting the background normalization, and thus directly translates
to large uncertainties in the $W$ tagging efficiency measurement. 
By using the $\tau_{21}'$, a significant improvement
is observed. 

To demonstrate this, we examine two cases, modified mass drop tagging with 
$\tau_{21} < 0.45$, and modified mass drop tagging with a
scale-dependent selection $\tau_{21} < 0.6 - 0.08 \times \rho'$, where 
$\rho' = \log{\left( m^2/p_T/\mu \right)}$.
This translates into a cut on $\tau_{21}' < 0.6$. These selections have approximately
the same signal efficiency. For simplicity, the same signal
and background MC samples are used as in the previous sections, but
the events are weighted with an easily specifiable fraction of background
jets. 
In this case, the background fraction for the
entire sample is 40\%. This gives a comparable fraction of merged to
unmerged $W$ bosons in a semileptonic $t\bar{t}$ selection at 13 TeV at the LHC,
but allows us to easily tune the fraction. 
In addition, to mimic the approximate detector resolution, the intrinsic resolution of the $W\rightarrow qq$ system
is smeared with a Gaussian of width 10 GeV. This is indicative of the resolutions obtained
at the CMS and ATLAS experiments. 

\begin{figure}[thb]
  \centering
  \includegraphics[scale=0.38]{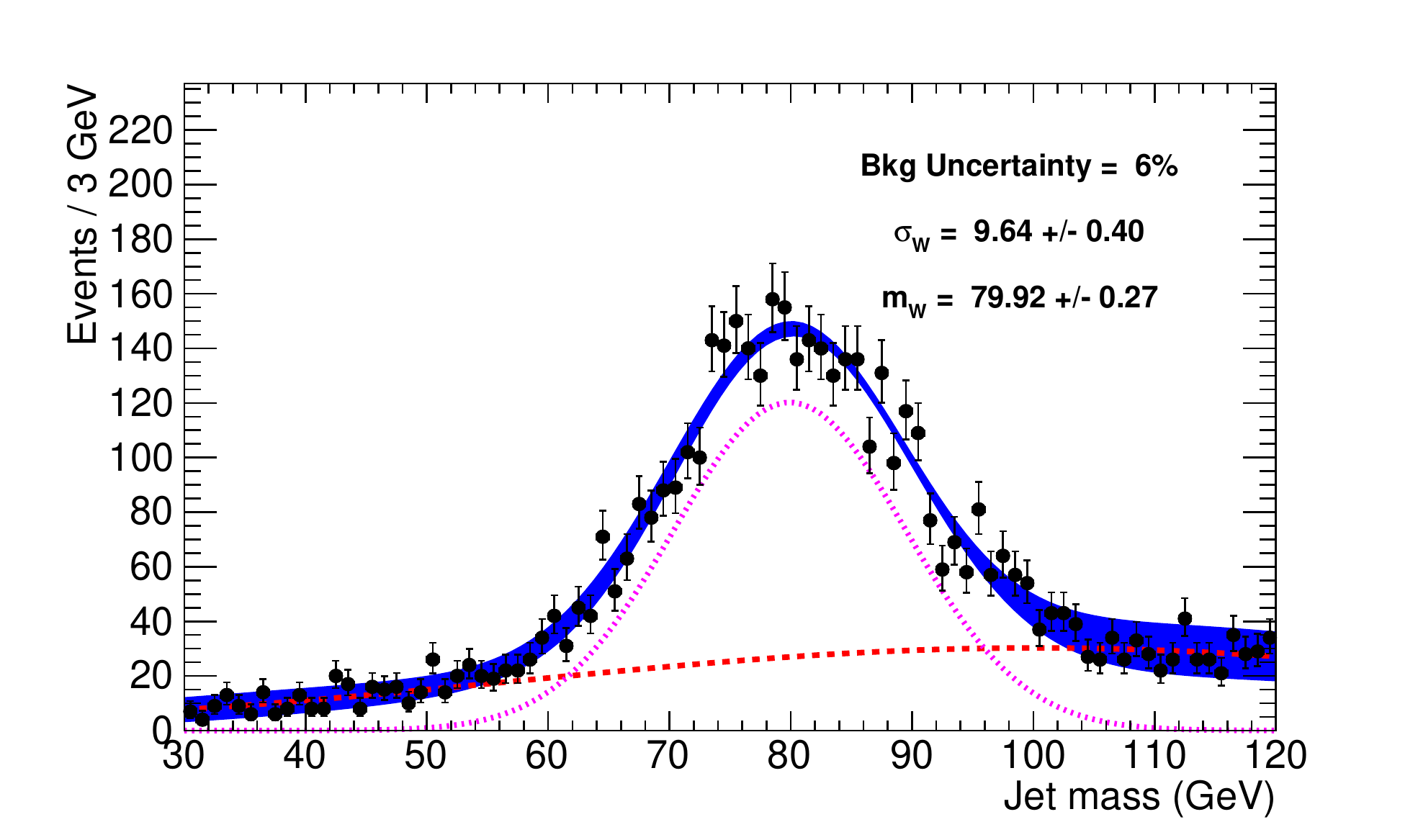}
  \includegraphics[scale=0.38]{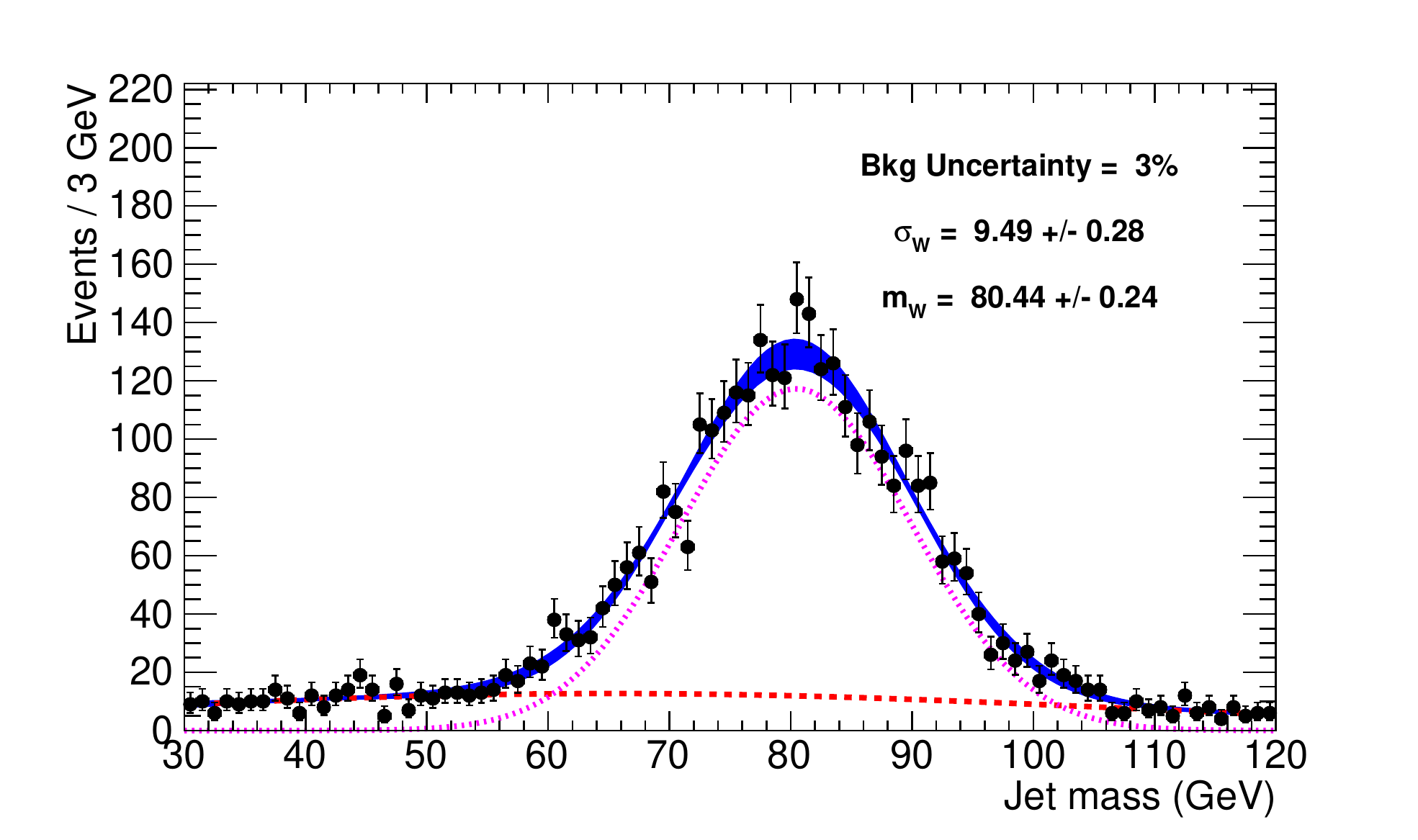}
  \caption{Jet mass for jets with \pt $=300-400$ GeV for     
    the modified mass-drop tagger after requiring $\tau_{21} < 0.45$ (left) and $\tau_{21}' < 0.6$ (right), respectively. 
    These two selections have approximately the same signal efficiency. The
    background fraction of the entire sample (for all jet masses) is set to 40\%. The points are the
    observed MC events, after smearing the jet mass resolution to $\sim 10\%$. 
    The purple dotted line corresponds to the smeared $W$ signal jets. The red dashed line 
    corresponds to the fitted background component, modeled as a 
    Gaussian distribution. The blue band corresponds to a fit to the signal plus background, where the thickness
    of the line corresponds to the uncertainty in the fitted component. }
  \label{fig:compare_fits_highpur_pt0300-0400}
\end{figure}

Figure~\ref{fig:compare_fits_highpur_pt0300-0400} shows simple fits to the 
jet mass for 5000 MC events in the range $50 < m_{J} < 120$ GeV, after a selection on the $N$-subjettiness variable. 
The model is a double Gaussian, one for the QCD continuum and one for the $W$ mass peak.
The jet \pt range 
considered is \pt $= 300-400$ GeV, to give a typical \pt range of the $W$ bosons
from top quark decays from SM $t\bar{t}$ production. The first fit
shows the modified mass drop algorithm after $\tau_{21} < 0.45$. The second
fit shows the modified mass drop algorithm after $\tau_{21}' < 0.6$. 
The fits successfully capture the mass of the $W$ and the input width of 10\%. 

It is interesting to note that the jet mass of the QCD jets after the $\tau_{21}'$ selection
are significantly pushed below 10 GeV. In addition, the remaining distribution is flat.
However, for the standard $\tau_{21}$ selection, the distribution is rising, with significantly
more background under the $W$ signal peak. 

The background uncertainty on the fit is is 6\% when using the standard $\tau_{21}$ selection.
However, it is reduced by a factor of two to 3\% by using the $\tau_{21}'$ selection. This is driven
by the fact that the fitter can more easily handle sidebands that are 
flatter, so the $\tau_{21}'$ variable outperforms the $\tau_{21}$ variable in
this metric. 

This would translate directly into a decreased systematic uncertainty for the LHC experiments.
While newer and more clever algorithms can achieve better performance in MC simulations, this does not
always translate directly to improvements in actual analyses due to the need to characterize
the systematic uncertainties. We therefore propose this test as an appropriate metric to characterize 
the systematic performance of new substructure algorithms. 

\subsection{Diboson Background Estimate}

The diboson background estimate for the LHC experiments is much the
same as the extraction of the $W$ tagging efficiency, except that the
background fraction is significantly higher. We have chosen a value of
80\% (integrated over the entire spectrum of events) as an indicative
fraction, with the same number of events (5000). We have considered two different \pt ranges, 
\pt $=500-600$ GeV and \pt $=1000-1100$ GeV. 

One somewhat obvious but important point is that as the \pt increases, the Sudakov peak
from QCD-generated jets shifts further to the right. As this occurs, the 
fits to discriminate boosted $W$ bosons from QCD-generated jets are less and
less able to distinguish between the categories.

\begin{figure}[thb]
  \centering
  \includegraphics[scale=0.38]{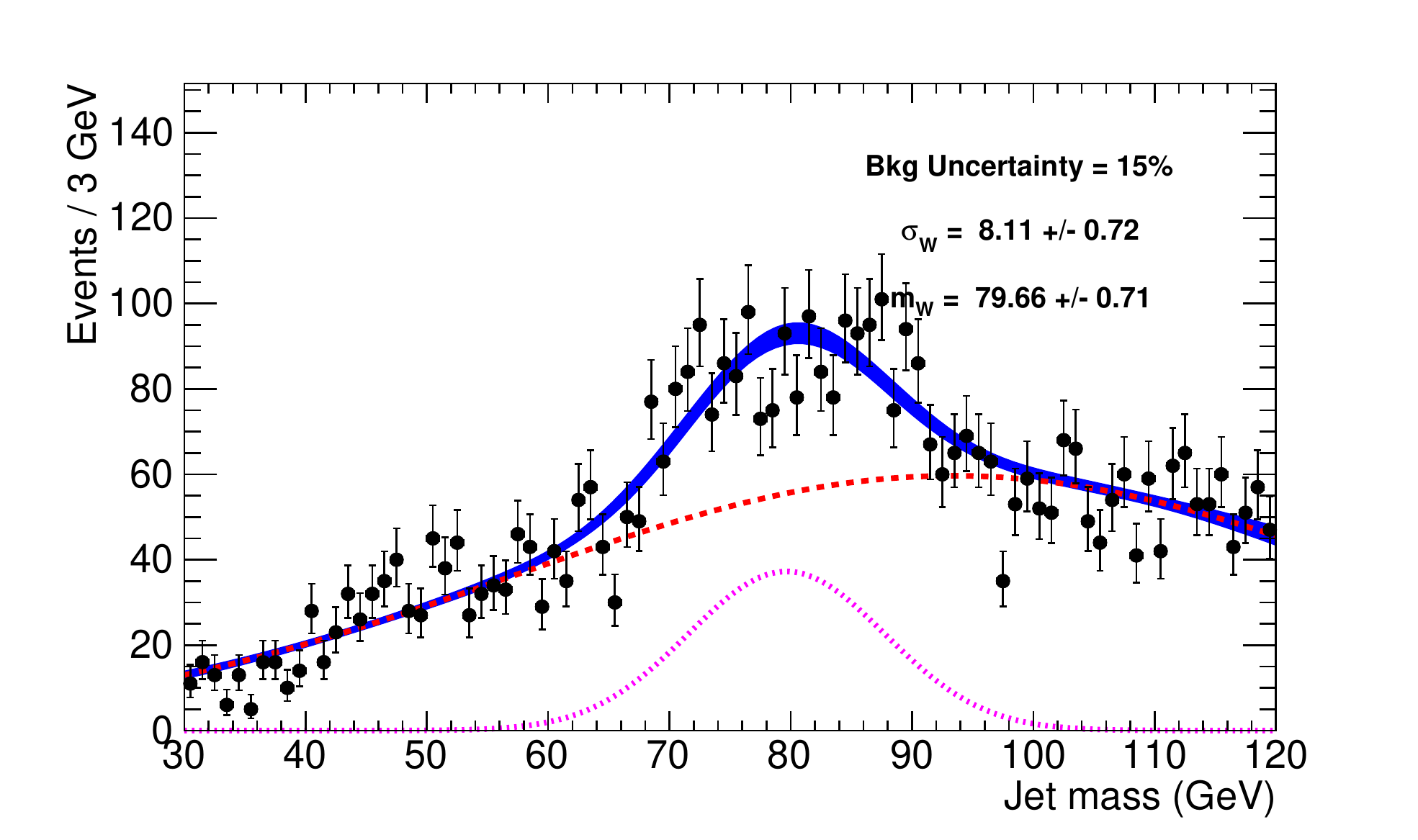}
  \includegraphics[scale=0.38]{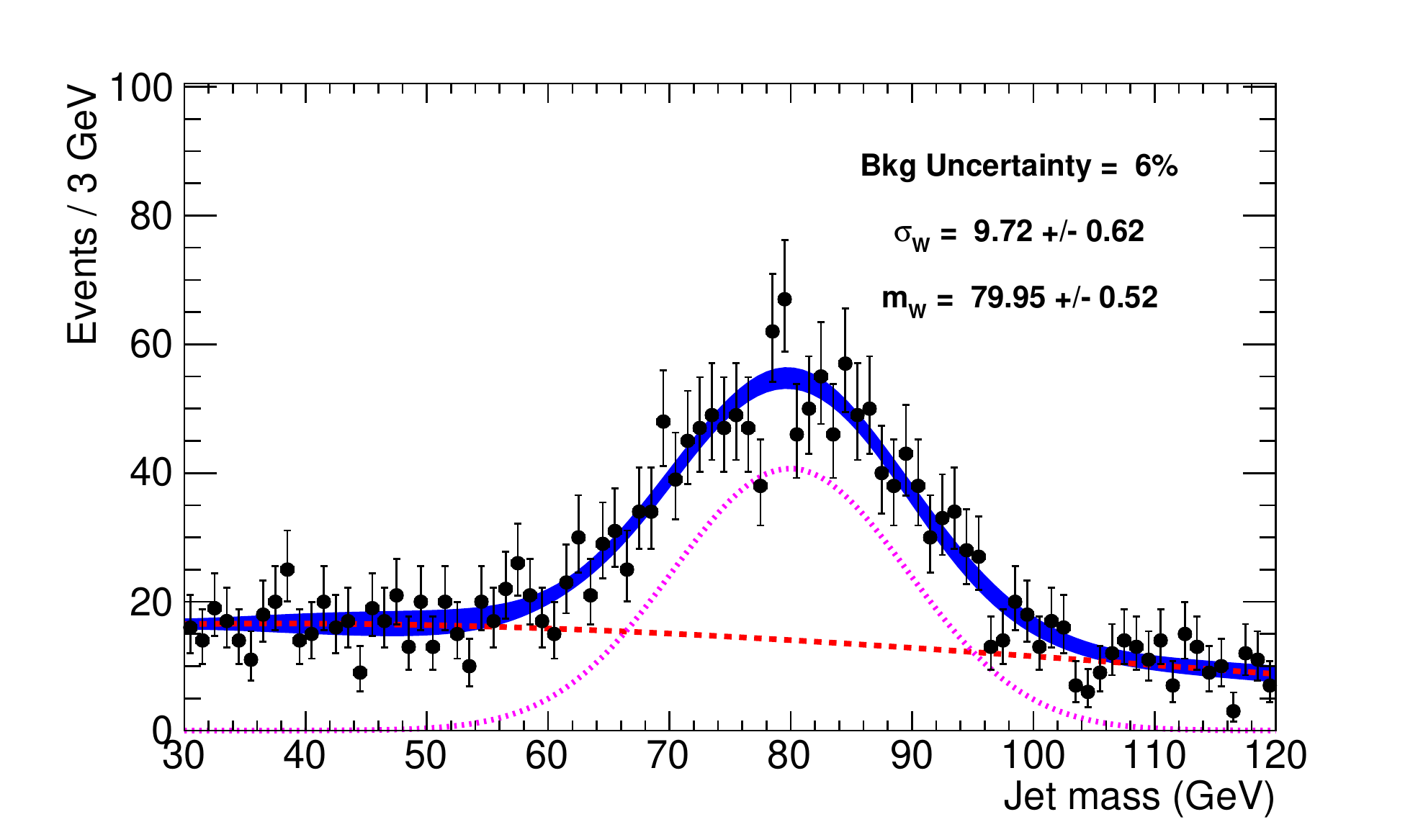}
  \caption{Jet mass for jets with \pt $=500-600$ GeV for 
    the modified mass-drop tagger after requiring $\tau_{21} < 0.45$ and $\tau_{21}' < 0.6$, respectively. 
    These two selections have approximately the same signal efficiency. The
    background fraction of the entire sample (for all jet masses) is set to 80\%. The points are the
    observed MC events, after smearing the jet mass resolution to $\sim 10\%$. 
    The purple dotted line corresponds to the smeared $W$ signal jets. The red dashed line 
    corresponds to the fitted background component, modeled as a 
    Gaussian distribution. The blue band corresponds to a fit to the signal plus background, where the thickness
    of the line corresponds to the uncertainty in the fitted component.}
  \label{fig:compare_fits_pt0500-0600}
\end{figure}

\begin{figure}[thb]
  \centering
  \includegraphics[scale=0.38]{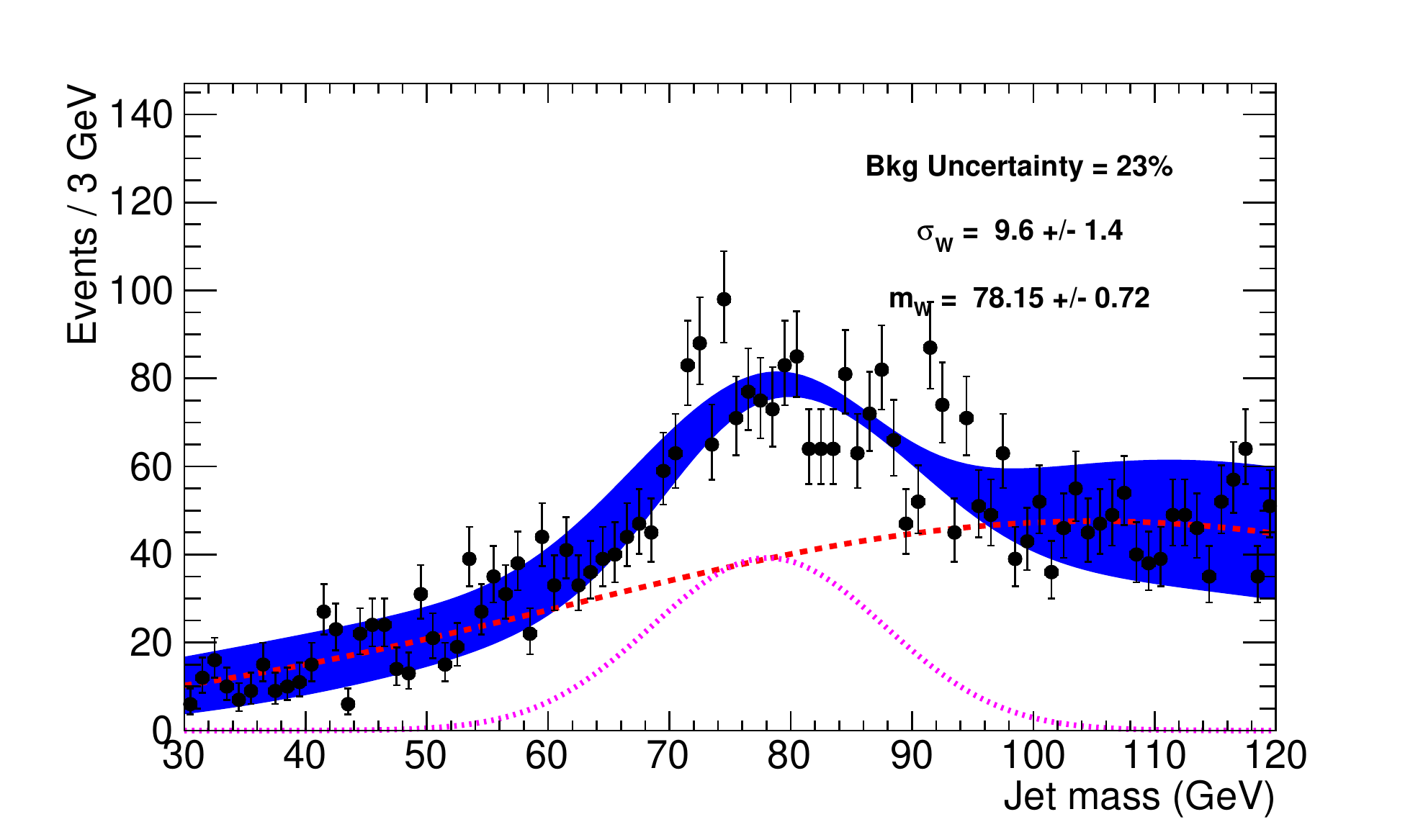}
  \includegraphics[scale=0.38]{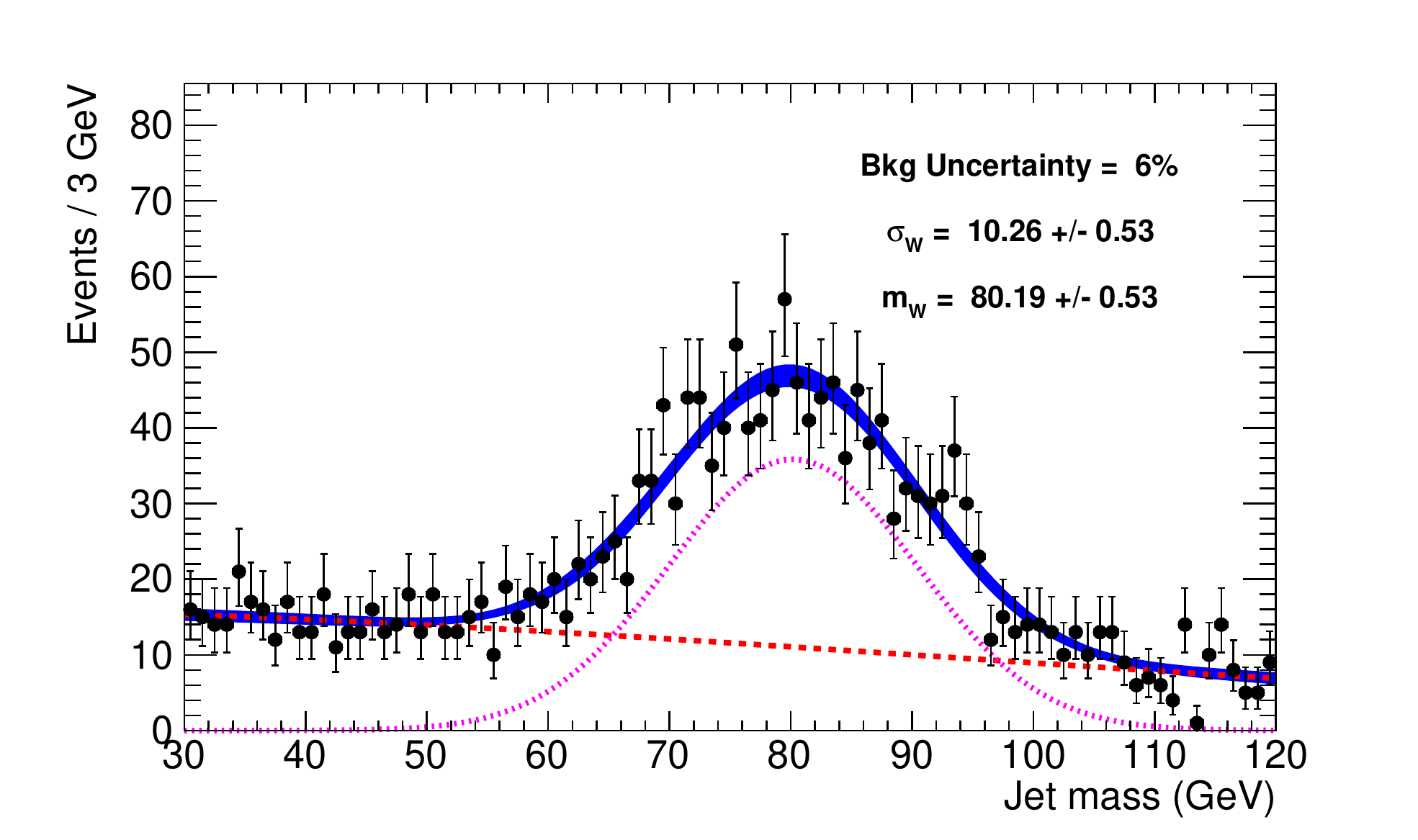}
  \caption{Jet mass for jets with \pt $=1000-1100$ GeV for 
    the modified mass-drop tagger after requiring $\tau_{21} < 0.45$ and $\tau_{21}' < 0.6$, respectively. 
    These two selections have approximately the same signal efficiency. The
    background fraction of the entire sample (for all jet masses) is set to 80\%. The points are the
    observed MC events, after smearing the jet mass resolution to $\sim 10\%$. 
    The purple dotted line corresponds to the smeared $W$ signal jets. The red dashed line 
    corresponds to the fitted background component, modeled as a 
    Gaussian distribution. The blue band corresponds to a fit to the signal plus background, where the thickness
    of the line corresponds to the uncertainty in the fitted component.}
  \label{fig:compare_fits_pt1000-1100}
\end{figure}

Figures~\ref{fig:compare_fits_pt0500-0600} and~\ref{fig:compare_fits_pt1000-1100} show 
similar fits as shown in Fig.~\ref{fig:compare_fits_highpur_pt0300-0400}. 
However, the background fraction is raised from 40\% to 80\% (again integrated over
the entire mass spectrum), and the \pt ranges are set to \pt $=500-600$ GeV
and \pt $=1000-1100$ GeV, respectively. 

For the range \pt $=500-600$ GeV, 
it is plain to see that there is a significant improvement of the 
$\tau_{21}'$ variable, where the background uncertainty decreases from 15\% to 6\%. This is even more
apparent for the range \pt $= 1000-1100$ GeV,where the
uncertainty decreases from 23\% to 6\%.

\section{Generalized Scale Invariance}
\label{ref:sec:generalized}
Decorrelation schemes can be extended beyond a pair of variables to
decorrelate classes of many variables. Such a procedure can be used to
allow for a class of variables to be merged into a single
multi-variate analysis discriminator (MVA), while preserving decorrelation against
one or a set of variables that are further used in the analysis.
Consider, for example, building an MVA $W$ tagger using both
$\tau_{2}/\tau_{1}$ and $C_{2}^{\beta=1}$. Both of these variables have
correlations with \pt and mass, so the resulting classifier that
combines the variables will also be correlated with mass and \pt.  Decorrelating the space of
variables against mass and \pt before or during the construction of
the MVA can thus preserve the mass and \pt invariance resulting in an uncorrelated
tagger. This idea has previously been pursued in $b$-physics utilizing
an MVA that minimizes the mass dependence, while simultaneously
constructing a classifier~\cite{Rogozhnikov:2014zea}. 

In light of building an example based on previously presented studies,
we split $\rho=\log(m^2/p_{T}^2)$ by into it components $\log(m)$ and
$\log(p_T)$. Combining this with either $C^{1}_{2}$ or $\tau_{2}/\tau_{1}$
gives a class of three variables for which we decorrelate into a set
of three independent linear combinations of variables. The independent variables can be
viewed as properties of the data which span the space of distinctive features.
This space can be explored to further understand behavior of the
data. Additionally, a subset of the independent components can be merged through an MVA
while maintaining the decorrelation of the remaining set of
variables. In this way, mass sidebands or other sideband methods
can be used on the merged MVA discriminator with the decorrelated
variable. 

As has previously been noted, decorrelating variables which are not implicitly linearly correlated is
poorly defined~\cite{Larkoski:2014pca}. We thus consider two generalized approaches that attempt to decorrelate
discriminators that are not necessarily linearly correlated. We
consider two decorrelation approaches: Principle Component Analysis (PCA) of transformed variables
and Independent Component Analysis (ICA). 

\subsection*{Decorrelation by PCA and ICA }
Given a set of variables  need not be linearly correlated, we consider
a transformed variable ($v'_{i}$) of the original variable $v_{i}$ defined by 
\begin{eqnarray}
v'_{i} & = & f(v_{i})
\end{eqnarray}
For this transformation, we train a gradient boosted decision
tree~\cite{Hocker:2007ht} with the boosted $W$ boson as a signal and a
high \pt QCD jet as a background. This transformation places the
variables into a space that enables the possibility of
linearized correlations of the original variables. 

The resulting correlation matrix of the transformed variables can be decorrelated through principle
component analysis by taking the eigenvectors of the matrix. This yields a set of $n$-independent vectors for a $n$-dimensional
correlation matrix. 

\begin{figure}[thb]
  \centering
  \includegraphics[scale=0.38]{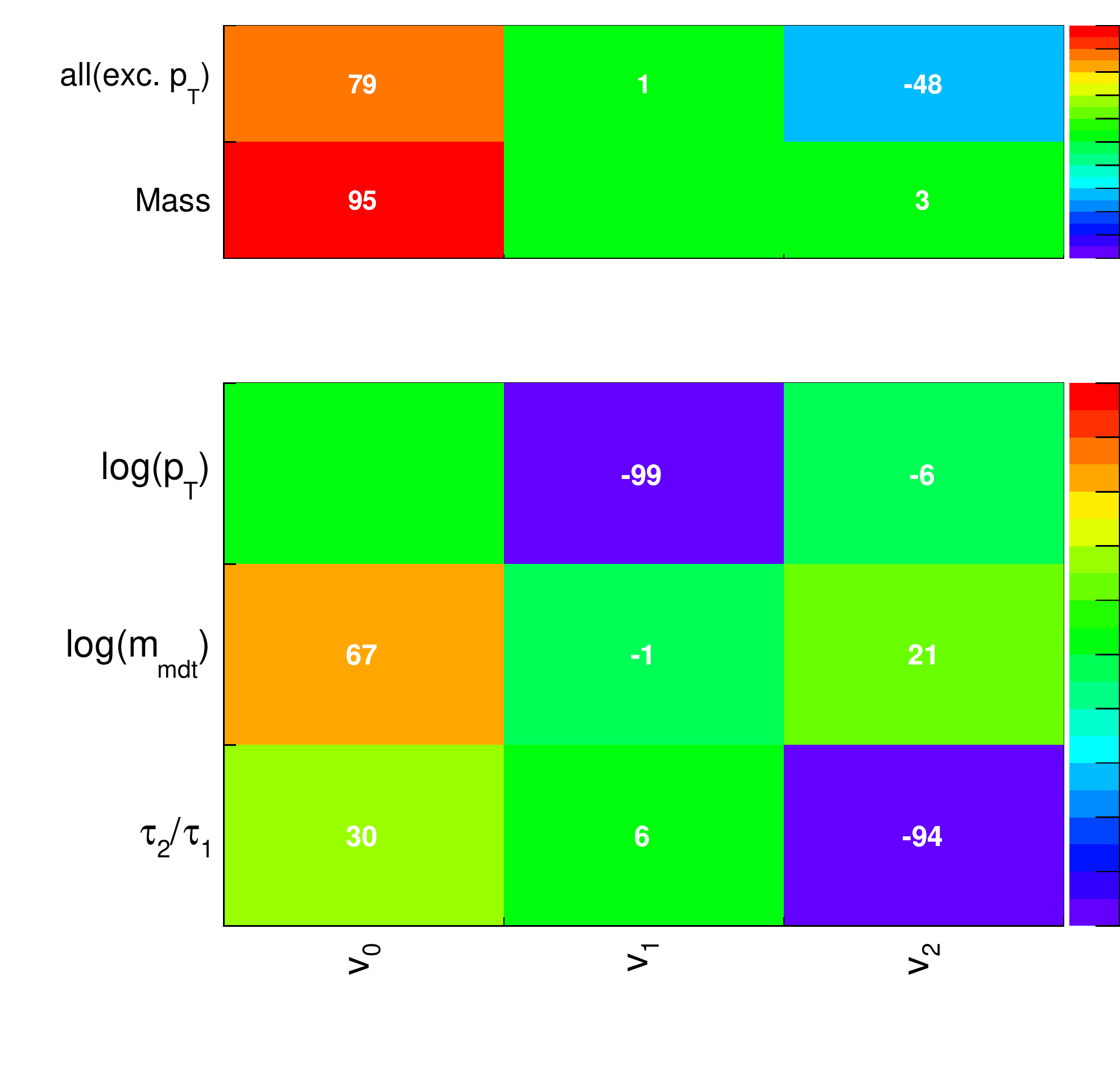}
  \includegraphics[scale=0.38]{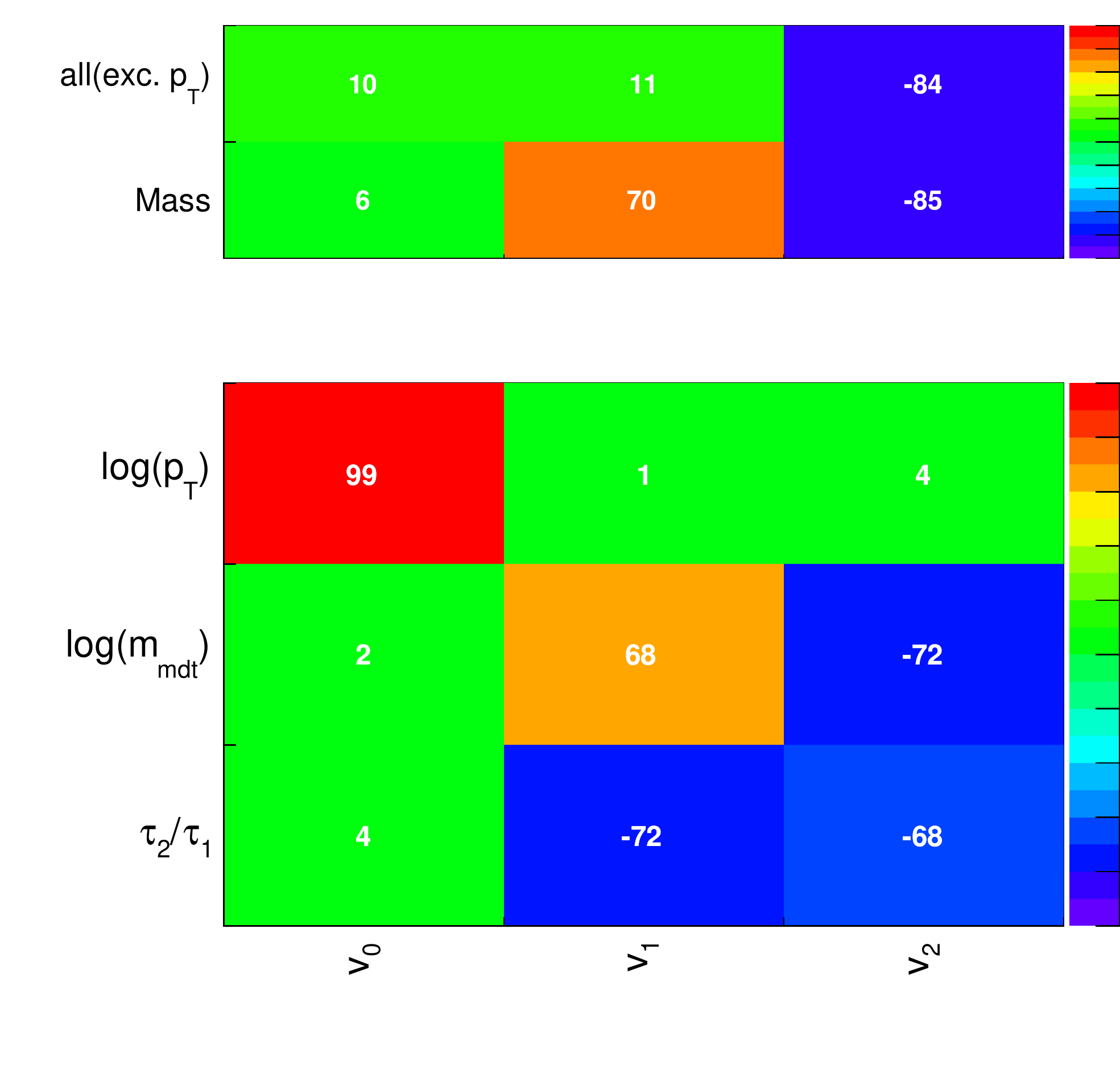}
  \caption{Decorrelated vectors from variable transformed PCA (left)
    and ICA (right). The bottom panel corresponds to the
    vectors in columns with their relative fraction labeled by
    row. The top panel corresponds to the correlation to the soft
    dropped mass and a gradient boosted decision tree trained with
    all variables excluding the \pt.  }
  \label{fig:t2t1ca}
\end{figure}

The decorrelated vectors for the triplet of transformed
$\tau_{2}/\tau_{1}$, $\log(p_{T})$, and $\log(m)$ is shown in
Fig.~\ref{fig:t2t1ca}. The correlation of the resulting vectors is
compared with a gradient boosted decision tree using all variables
and with the transformed mass. From this correlation,  we observe two discriminating
dimensions and the $p_{T}$. These we can write as 
\begin{eqnarray}
v_{1} & = & \log(m/\mu_{1})+ K_{1}(\tau_{2}/\tau_{1}) \\
v_{2} & = & \tau_{2}/\tau_{1} + K_{2}  \log(m^{3.5}/p_{T}\mu^{2.5}_{2}), 
\end{eqnarray}
where $K_{1,2}$ correspond to coefficients and $\mu_{1,2}$ are scales, typically $\mu_{1,2}\sim1$~GeV  to make the observables dimensionless. The first variable corresponds to the transformed mass and the second
corresponds the transformed $\tau_{2}/\tau_{1}$. The second variable
is not too different from $\rho^{\prime}$ decorrelated
$\tau_{2}/\tau_{1}$. 

An alternative decorrelation approach, known as independent
component analysis (ICA), involves diagonalization of the matrix
constructed by computing the pairwise mutual information of each pair
of variables on the sample of QCD jets. This differs to previous
approaches, which rely on the mutual information to truth. Here, we
focus on identifying features in the data and not necessarily
discriminating power. We perform the ICA with an algorithm
that uses $k$-nearest neighbor to expedite the diagonalization
process (MILCA)~\cite{kraskov:2004}. The right panel of Fig.~\ref{fig:t2t1ca} shows the ICA decomposed
vectors. As with the transformed PCA, the ICA decorrelates the \pt,
however the mass $\tau_{2}/\tau_{1}$ interdependence is stronger than
in the transformed case. 

\begin{figure}[thb]
  \centering
  \includegraphics[scale=0.50]{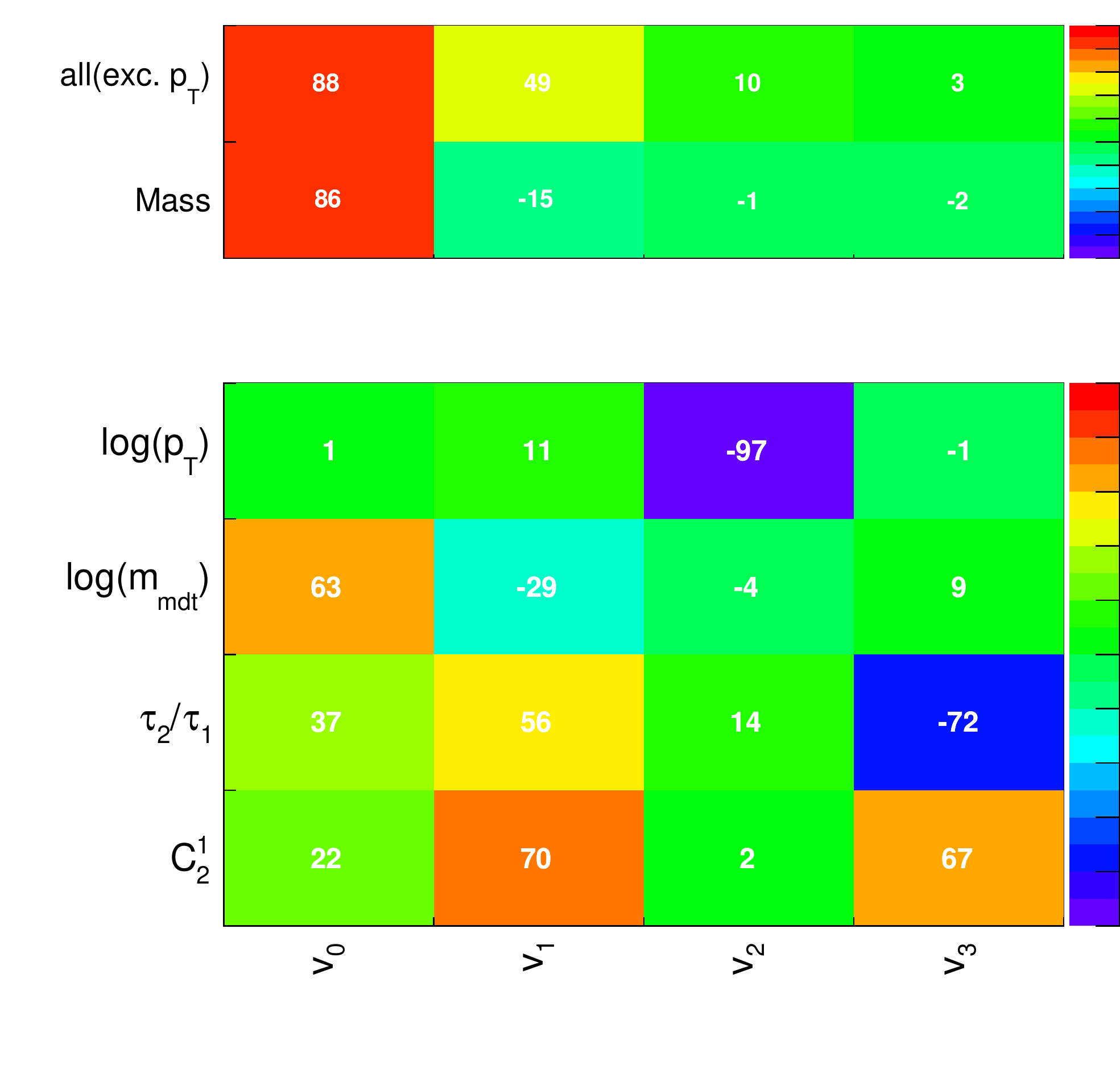}
  \caption{Decorrelated vectors from variable transformed PCA Using
    the full suite of observables studied in this paper. The bottom panel corresponds to the
    vectors in columns with their relative fraction labeled by
    row. The top panel corresponds to the correlation to the soft
    dropped mass and a gradient boosted decision tree trained with
    all variables excluding the \pt.  }
  \label{fig:ca}
\end{figure}

Finally, the equivalent decorrelated matrix for a combined
set of observables is shown in Fig.~\ref{fig:ca}, here we show just the
transformed PCA approach. From the combined
set, we observe the largest orthogonal set of discrimination power
comes from the $C_{2}^{\beta=1}$ as oppose to $\tau_{2}/\tau_{1}$.  
When comparing the two approaches, we have found variable transformed PCA yields a more consistent performance with our previous observations.

\section{Conclusion and Outlook}

In this note, we explore the scale-dependence and correlations of jet substructure observables.
The goal is not only to improve the statistical power of such observables, which we also demonstrate, but also to consider 
practical issues related to using such observables in searches for new physics.  
In order to design decorrelated taggers (DDT), we transform the shape observable, here $\tau_2/\tau_1 \rightarrow \tau_{21}'$, by decorrelating it from groomed mass observables also factoring in the $p_T$ scale-dependence.
In addition to improving the statistical discrimination between signal and background, we also preserve a robust, flat background shape and 
which has more stable behavior when scaling of the background going from lower $p_T$ bins to higher $p_T$ bins. 
We demonstrate the advantages of such an approach in various case studies such as predicting background normalizations and determining heavy object tagging scale factors related to new physics searches.

The intention of this note is not to perform a detailed study of all possible heavy object taggers, 
but instead, to introduce further considerations when designing taggers and propose a method by which all considerations can be addressed, namely via 
observable decorrelation.
We leave studies related to variations on jet mass groomers and shape observables, R-scaling, quark-gluon fractions, scaling background predictions, behavior at extremely high $p_T$,
and top tagging to future works.  
We have explored more generic determinations of observable decorrelation with complex taggers 
using multivariate techniques and numerical principle-component analysis.

\section*{Acknowledgments}
We thank Andrew Larkoski and Petar Maksimovic for their critical reading of the manuscript.
The authors would also like to thank Matteo Cremonesi, Matthew Low, 
 Cristina Mantilla Suarez, Siddharth Narayanan, Gavin Salam, Gregory Soyez, and Michael Spannowsky for useful discussion and inputs. 
JD and SR are partially supported by the U.S.\ National Science Foundation, under grant PHY--1401223.  
PH is supported by CERN.
The work of SM is partly supported by the U.S.\ National Science Foundation, under grant PHY--0969510, the LHC Theory Initiative.
NT is supported by the Fermi Research Alliance, LLC under Contract No. De-AC02-07CH11359 with the United States Department of Energy.

\bibliographystyle{jhep}
\bibliography{scot}

\end{document}